\documentclass[
preprint,showpacs,
amsmath,amssymb,aps
]{revtex4-1}
 
\usepackage{graphicx}
\usepackage{dcolumn}
\usepackage{bm}
\newcommand{\md}{\mathrm d}

\begin{document}

\title{Numerical solution for the Schwinger--Dyson equation\\
at finite temperature in Abelian gauge theory}

\author{Hiroaki Kohyama}
\affiliation{Department of Physics,
National Taiwan University, Taipei 10617, Taiwan.
}

\date{\today}

\begin{abstract}
We present the exact numerical solutions for the Schwinger--Dyson
equations at finite temperature with general gauge in Abelian gauge
theory. We then study the chiral phase transition on temperature
from the obtained solutions. We find that, within the quenched
Schwinger--Dyson equations, there exists substantial gauge
dependence on the solutions and the critical temperature.
\end{abstract}


\maketitle

\section{\label{sec:intro}
Introduction}
The chiral phase transition at finite temperature is an important
and interesting phenomenon in gauge theories. The chiral symmetry
is broken for low temperature in the strongly coupling system,  and
expected to be restored for high temperature. Then, it is interesting
to investigate this critical behavior; in particular, the critical
temperature on the chiral phase transition.

The Schwinger--Dyson equation (SDE)~\cite{Dyson:1949ha} is
frequently used method for the studies on the above mentioned
chiral phase transition. The SDE are the integral equations for the
green functions, and the solutions lead the field strength
renormalization and the dynamically generated mass (for reviews,
see,~\cite{Roberts:1994dr, Roberts:2000aa, Holl:2006ni}).
A lot of works have been devoted to the analyses on the SDE at
zero temperature~\cite{Johnson:1964da, Maskawa:1974vs,
FK76, Ball:1980ay, Roberts:1989mj, Curtis:1990zs, Bando:1993qy,
Kizilersu:2009kg, Kizilersu:2014ela}
and finite temperature~\cite{Ikeda:2001vc, Akiba:1987dr,
Kalashnikov:1987td, Harada:1998zq, Fukazawa:1999aj, 
Fueki:2002mm, Mueller:2010ah, Nakkagawa:2011ci}.
The investigations at finite temperature are rather limited comparing
to the ones at zero temperature, since it becomes difficult to solve the
equations at finite temperature due to the separation of the time and
space components~\cite{LeBellac}. Then the preceding works
have mostly been done through employing some approximations on
the equations in a specific gauge. However, it is possible to evaluate
the equations without approximations~\cite{Ikeda:2001vc}. In this
paper, we shall solve the SDE at finite temperature without
approximations in general gauge. Thereafter we study the gauge
dependence on the chiral phase transition with respect to temperature.

The plan of the paper is as follows. Section~\ref{sec:model}
presents the equations at finite temperature and the prescription of
solving the equations. The numerical solutions and the critical behavior
are shown in Secs.~\ref{sec:re_ns} and \ref{sec:re_cr}. The concluding
remarks are given in Sec.~\ref{sec:conclusion}.

\section{\label{sec:model}
Schwinger--Dyson equation}
We will here present the SDE at finite temperature and the iteration
method which is the numerical procedure employed in this letter.

\subsection{\label{subsec:equation}
The equations at finite temperature}
The equation for the fermion self-energy $\Sigma(p)$ is given by 
\begin{equation}
  \Sigma (p)
  = i e^2 \int \frac{{\mathrm d}^4 q}{(2\pi)^4}
     \gamma^{\mu} D_{\mu\nu}(p-q) S(q) \Gamma^{\nu}(p,q) ,
\end{equation}
with  the coupling strength, $e$, the propagators for the photon and
fermion, $D_{\mu \nu}(p-q)$ and $S(q)$, and the vertex function
$\Gamma^\nu (p,q)$. For $D_{\mu \nu}$ and $\Gamma^\nu$, we
use the following quenched forms
\begin{eqnarray}
  &&D_{\mu\nu}(k)
  =  \frac{-g_{\mu \nu}}{k^2}
    +(1-\xi) \frac{k_{\mu}k_{\nu}}{k^4},
  \label{eq:photon_prop} \\
  &&\Gamma^{\nu}(p,q) = \gamma^{\nu},
\end{eqnarray}
where $\xi$ is the gauge parameter and $k_\mu = p_\mu -q_\mu$.

The fermion propagator is defined by
\begin{equation}
  S(p_0, p) = \frac{1}
  { C(p_0,p)\gamma_0 p_0
    +A(p_0,p)\gamma_i p^i -B(p_0,p)}
\end{equation}
with $p = |{\bf p}|$. Note that the time and space components are to
be treated separately for the extension to finite temperature. 
In the imaginary time formalism, the continuous $q_0$ integral is
replaced by the discretized
summation due to the applied boundary condition~\cite{LeBellac},
\begin{equation}
  \frac{1}{2\pi i} \int_{-\infty}^{\infty} \md q_{0} F(q_0)
  \to
  T \sum_{m=-\infty}^{\infty} F(i \omega_m)
\end{equation}
where $T$ is the temperature of the system. The frequency, 
$\omega_m$, is taken as
\begin{equation}
  \omega_m = (2m + 1)\pi T
\end{equation}
for the fermionic field.

After some algebras one obtains the following equations
\begin{align}
 &C(\omega_n, p) = 1
   + \frac{\alpha}{\pi}
  T \sum_{m=-\infty}^{\infty}
  \int_{\lambda}^{\Lambda} \md q \,
     \left[ 
        {\mathcal I}_{CC} C(\omega_m, q) + {\mathcal I}_{CA} A(\omega_m, q)  
     \right]      \Delta(\omega_m, q) , \\
&A(\omega_n, p) = 1
   + \frac{\alpha}{\pi}
  T \sum_{m=-\infty}^{\infty}
  \int_{\lambda}^{\Lambda} \md q \,
     \left[
       {\mathcal I}_{AC} C(\omega_m, q) +  {\mathcal I}_{AA} A(\omega_m, q)  
     \right]      \Delta(\omega_m, q), \\
&B(\omega_n, p) = m_0
   +  \frac{\alpha}{\pi}
  T \sum_{m=-\infty}^{\infty}
  \int_{\lambda}^{\Lambda} \md q \,
     \left[ {\mathcal I}_{B} B(\omega_m, q)
     \right]      \Delta(\omega_m, q),
\end{align}
with $\alpha = e^2/(4\pi)$,
\begin{align}
  \Delta(\omega_m, q)
  =  \frac{1}{ C^2(\omega_m, q) \omega_m^2  
                 + A^2(\omega_m, q) q^2 
                 + B^2(\omega_m, q)}.
\end{align}
and
\begin{align}
  & {\mathcal I}_{CC} = 
     -\xi_{+} \frac{{\omega_m}}{\omega_n}  I_1
     +2 \xi_{-} \frac{{\omega_m}}{\omega_n} {\omega^{\prime}_m}^2 I_2, \\
  & {\mathcal I}_{CA} = 
       \xi_{-} \frac{\omega^{\prime}_m}{\omega_n}  I_1
     +\xi_{-} \frac{\omega^{\prime}_m}{\omega_n}  
        \left[ {\omega^{\prime}_m}^2 - q^2 + p^2 \right] I_2, \\
  & {\mathcal I}_{AC} = 
      -\xi_{-} \frac{{\omega_m} {\omega^{\prime}_m}}{p^2}  I_1
      -\xi_{-} \frac{{\omega_m} {\omega^{\prime}_m}}{p^2}  
        \left[ {\omega^{\prime}_m}^2 + q^2 - p^2 \right] I_2, \\             
  & {\mathcal I}_{AA} = 
      -\frac{2q^2}{p^2}  
      -\frac{1}{2p^2} \left[ \xi_{3}^{-} {\omega^{\prime}_m}^2 
        + \xi_+(q^2+p^2) \right] I_1
      -\frac{1}{2p^2} \xi_{-} 
        \left[ {\omega^{\prime}_m}^4 - (q^2-p^2)^2 \right] I_2, \\
   &  {\mathcal I}_{B} = -\xi_3^{+} I_1 , \\
   &I_1 = \frac{q}{2p}
         \ln \frac{{\omega^{\prime}_m}^2+(q-p)^2}
                       {{\omega^{\prime}_m}^2+(q+p)^2}, \\
  &I_2 = \frac{q}{2p}
        \left[
          \frac{1}{{\omega^{\prime}_m}^2+(q-p)^2}
          -\frac{1}{{\omega^{\prime}_m}^2+(q+p)^2}
        \right].
\end{align}
Here ${\omega^{\prime}_m} \equiv {\omega_n}-{\omega_m}$,
$\xi_{\pm} \equiv 1 \pm \xi$ and $\xi_3^{\pm} = 3 \pm \xi$,
and we introduced the infrared and ultraviolet cutoffs, $\lambda$ and
$\Lambda$. These equations are the set of the Schwinger--Dyson
equations at finite temperature which are the straightforward extension
from the equations at zero temperature with three dimensional
momentum cutoff~\cite{Kohyama:2015fta}.

Once the solutions are obtained, we can evaluate the dynamical mass
and the chiral condensate through the relations,
\begin{align}
  &M(\omega_n, p) = B(\omega_n, p)/A(\omega_n, p), \\
  & \phi = \langle \bar{\psi} \psi \rangle
  = -T \sum_{m=-\infty}^{\infty}
     \int \frac{\md^3 q}{(2\pi)^3} {\rm tr}
     \left[
         \frac{B(\omega_m, q) }{ C^2(\omega_m, q) \omega_m^2  
                 + A^2(\omega_m, q) q^2  + B^2(\omega_m, q)}
     \right] .
\end{align}
It should be noted that the chiral condensate becomes zero when the
dynamical mass vanishes. Therefore one can see whether the chiral
symmetry is broken, $\phi \neq 0$, from the value of $M(\omega_n, p)$.
Strictly, the phase transition should be evaluated at the minimum of the
effective potential which is derived by integrating out the gap equation
with respect to the order parameter. However,  it is known to be adequate
to see the solution of the gap equation in the system at zero chemical
potential~\cite{Harada:1998zq}, then we search the critical temperature
on the chiral phase transition by checking the value of the dynamical
mass in this paper.

\subsection{\label{subsec:iteration}
The iteration method}
In this subsection, we discuss the numerical procedure for the
integral equations. To briefly present the basic idea of the iteration
method, we show the equations with the trapezoidal rule. (In
the actual calculations, we used the Gauss-Legendre integration
for the $q-$integration.)

In solving the equations, we discretize the integral to discrete
summation as
\begin{align}
 &C_{nj}^{(l)} = 1
   + \frac{\alpha}{\pi}
  T \sum_{m=-N_t}^{N_t-1}
  \delta \sum_{k=0}^{N_p-1}
     \left[ 
       {\mathcal I}^{CC}_{njmk} C_{mk}^{(l-1)}
       + {\mathcal I}^{CA}_{njmk} A_{mk}^{(l-1)}
     \right]      \Delta_{mk}^{(l-1)} , \\
&A_{nj}^{(l)} = 1
   + \frac{\alpha}{\pi}
  T \sum_{m=-N_t}^{N_t-1}
  \delta \sum_{k=0}^{N_p-1}
     \left[
        {\mathcal I}^{AC}_{njmk} C_{mk}^{(l-1)}
      +{\mathcal I}^{AA}_{njmk} A_{mk}^{(l-1)}
     \right]      \Delta_{mk}^{(l-1)}, \\
&B_{nj}^{(l)} = m_0
   +  \frac{\alpha}{\pi}
  T \sum_{m=-N_t}^{N_t-1}
  \delta \sum_{k=0}^{N_p-1}
     \left[ {\mathcal I}^{B}_{njmk} B_{mk}^{(l-1)}
     \right]      \Delta_{mk}^{(l-1)},
\end{align}
with $\delta \equiv (\Lambda-\lambda)/N_p$. Note that we introduce
the cutoff on the frequency summation since it is technically impossible
to take the infinite number of summation. However, as will be confirmed
by the numerical calculations in the next section, the contributions
from the large number of $n$ and $m$ are negligible,
then the artifact of setting $N_t$ does not affect the solutions
if we choose large enough number for $N_t$. The matrix of the form
$F_{njmk}$ indicates
\begin{equation}
  F_{njmk} = F(\omega_n, p, \omega_m, q), \quad
  p = \lambda + \delta j, \quad
  q = \lambda + \delta k,
\end{equation}
and the superscript $l$ in $F^{(l)}$ means the number of iteration.
Therefore, the $l\,$th solutions, $F^{(l)}$, are derived from the equations
with the previous solutions, $F^{(l-1)}$. By setting the initial conditions
on $C^{(0)}$, $A^{(0)}$ and $B^{(0)}$ then carrying out the enough times
of iterations, one can obtain the solutions.

\section{\label{sec:re_ns}
Numerical solution}
Having presented the equations and the procedure of solving them,
we are now ready to perform the numerical analyses. In this section,
we are going to show the numerical solutions for $C$, $A$ and $B$
with respect to $\omega_n$, $p$, and temperature.

\subsection{\label{subsec:ABCnp}
Solutions for $C(\omega_n,p)$, $A(\omega_n,p)$, $B(\omega_n,p)$}
\begin{figure}[!h]
\begin{center}
  \includegraphics[width=5.3cm,keepaspectratio]{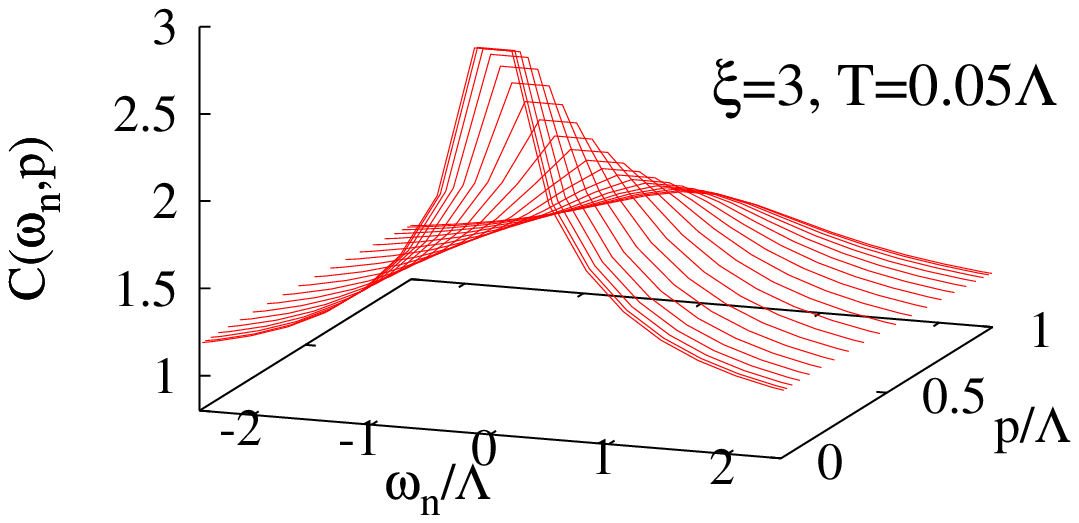}
  \includegraphics[width=5.3cm,keepaspectratio]{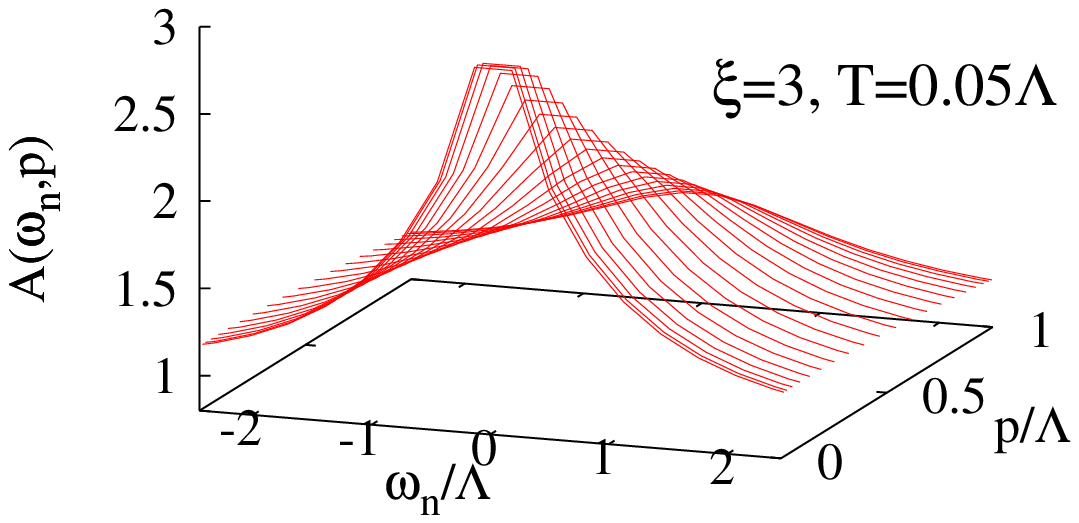}
  \includegraphics[width=5.3cm,keepaspectratio]{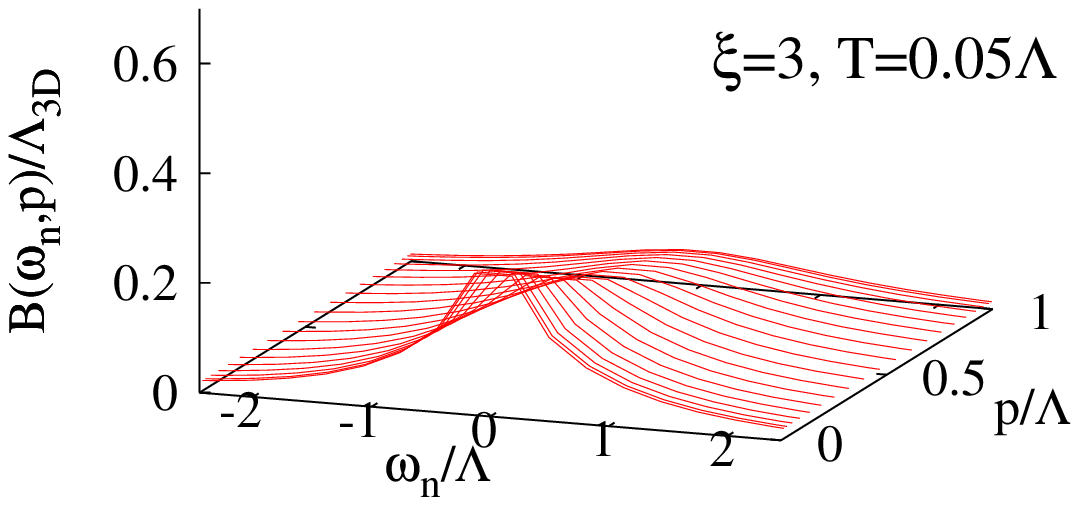}
  \includegraphics[width=5.3cm,keepaspectratio]{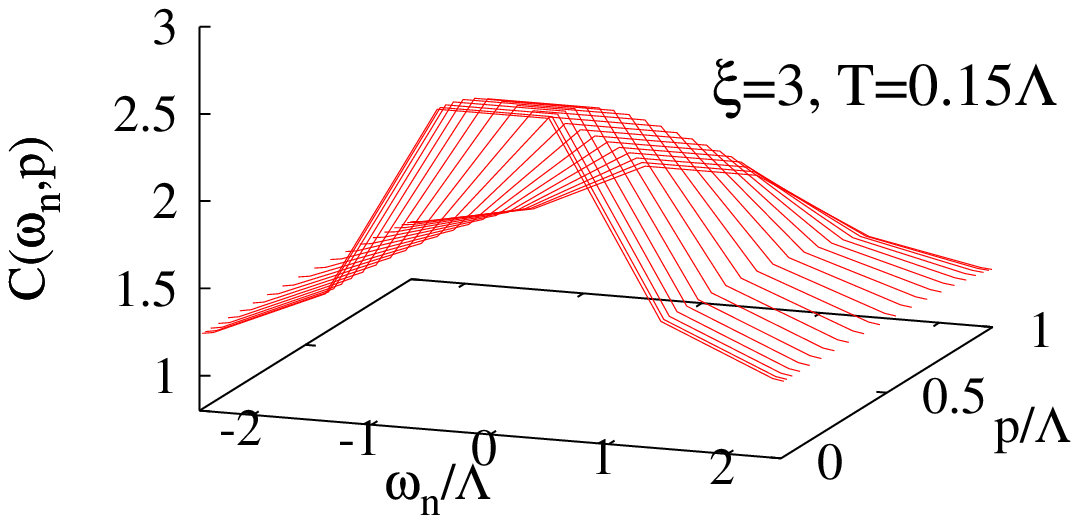}
  \includegraphics[width=5.3cm,keepaspectratio]{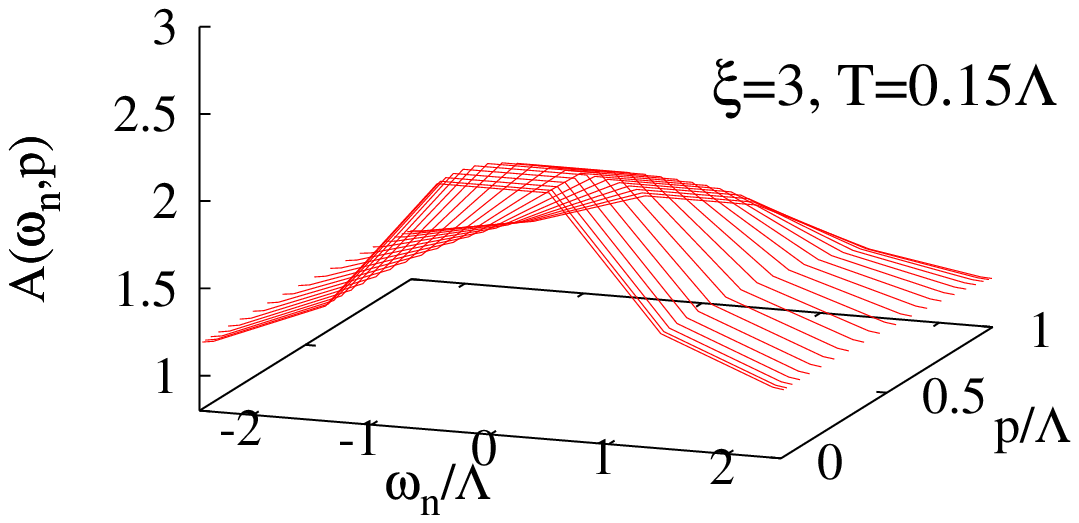}
  \includegraphics[width=5.3cm,keepaspectratio]{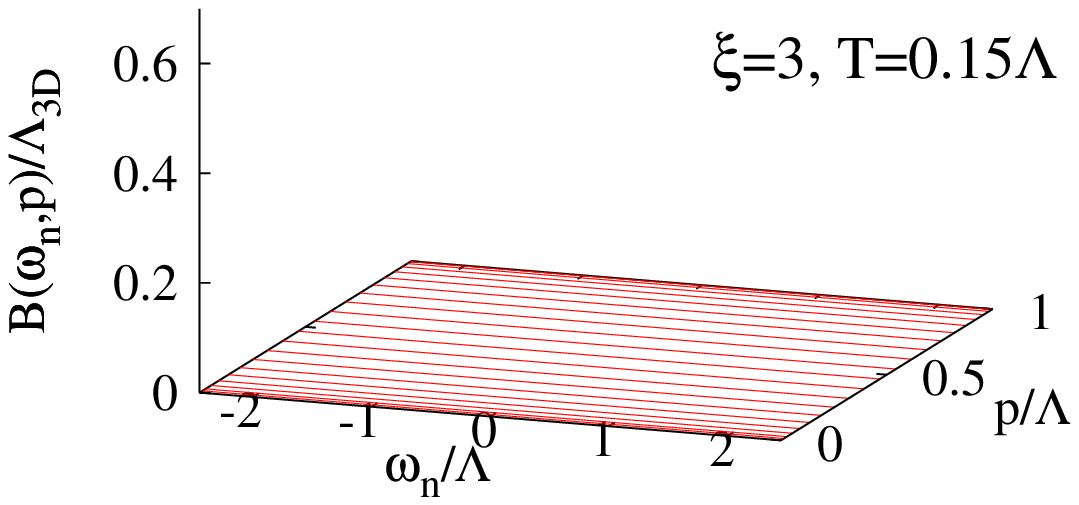}
  \includegraphics[width=5.3cm,keepaspectratio]{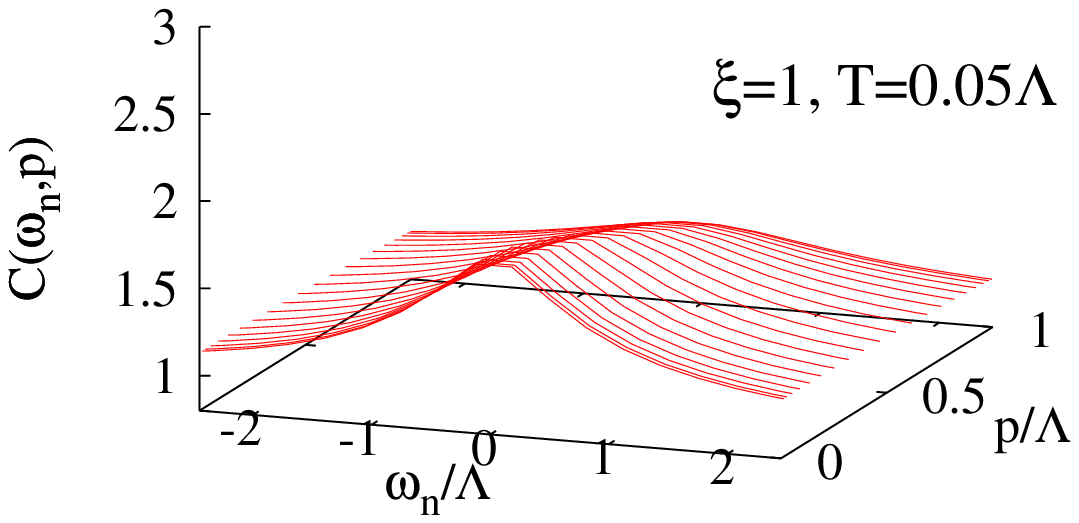}
  \includegraphics[width=5.3cm,keepaspectratio]{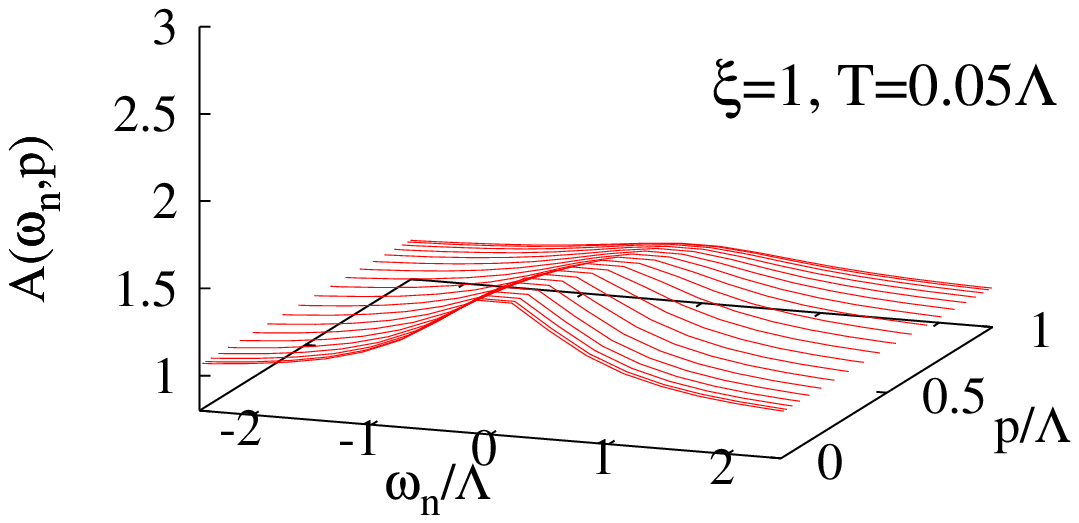}
  \includegraphics[width=5.3cm,keepaspectratio]{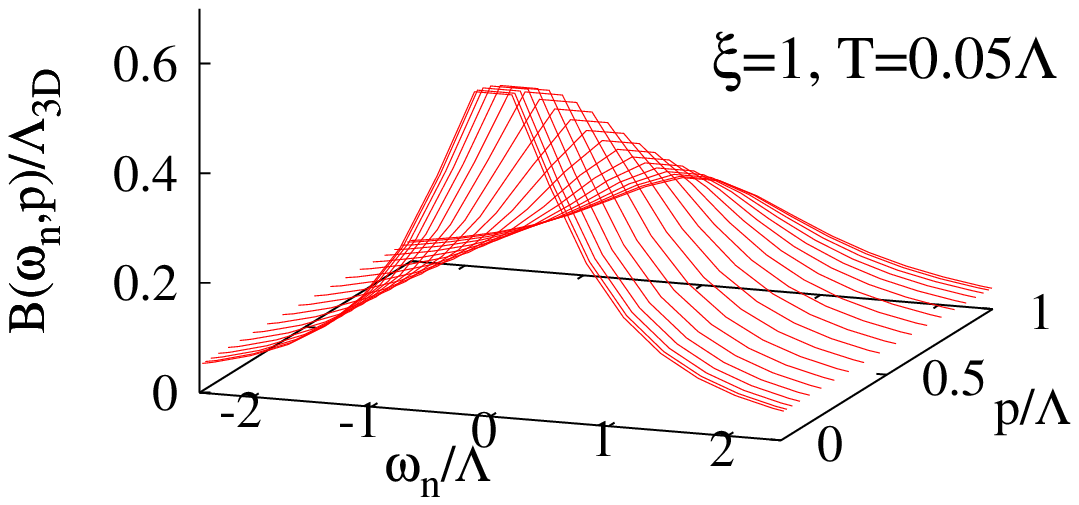}
  \includegraphics[width=5.3cm,keepaspectratio]{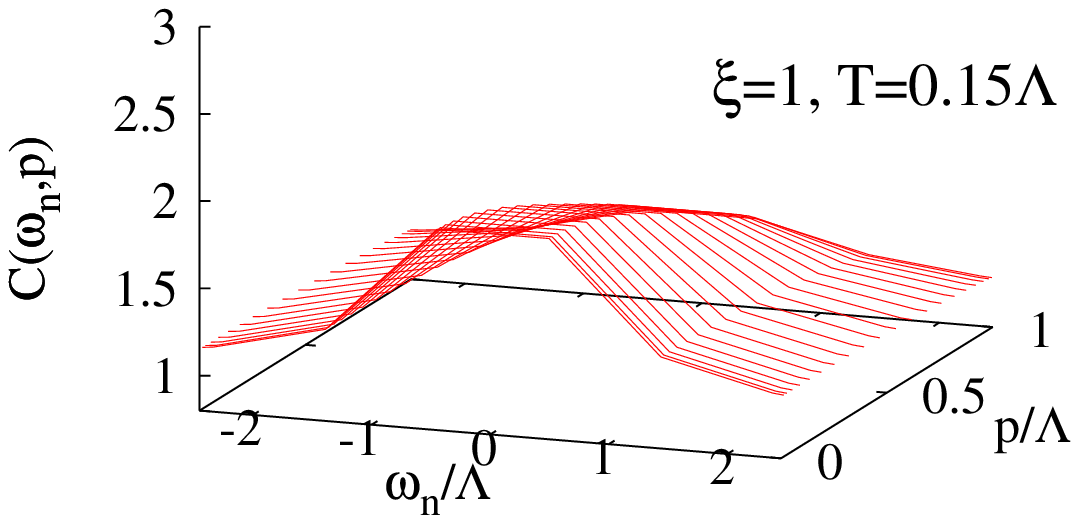}
  \includegraphics[width=5.3cm,keepaspectratio]{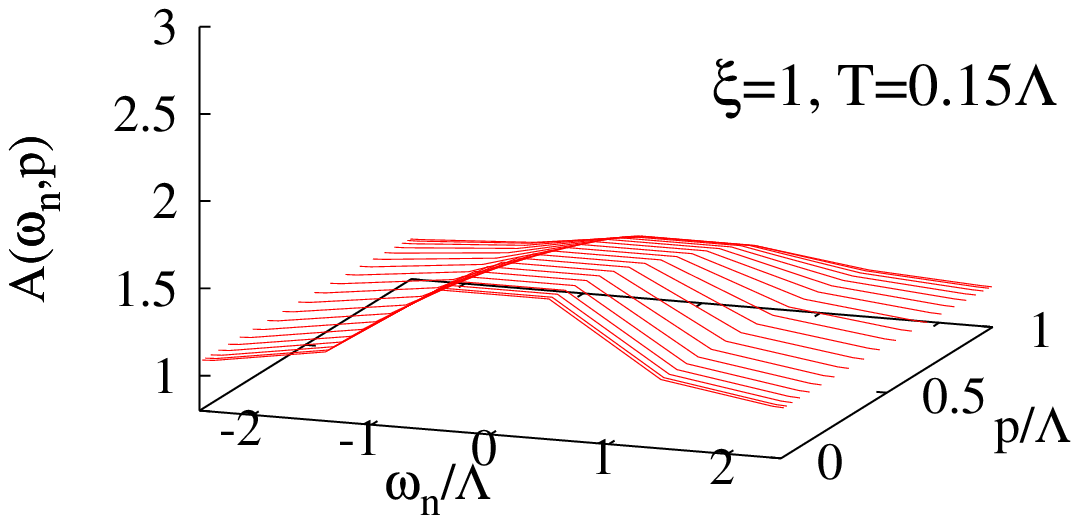}
  \includegraphics[width=5.3cm,keepaspectratio]{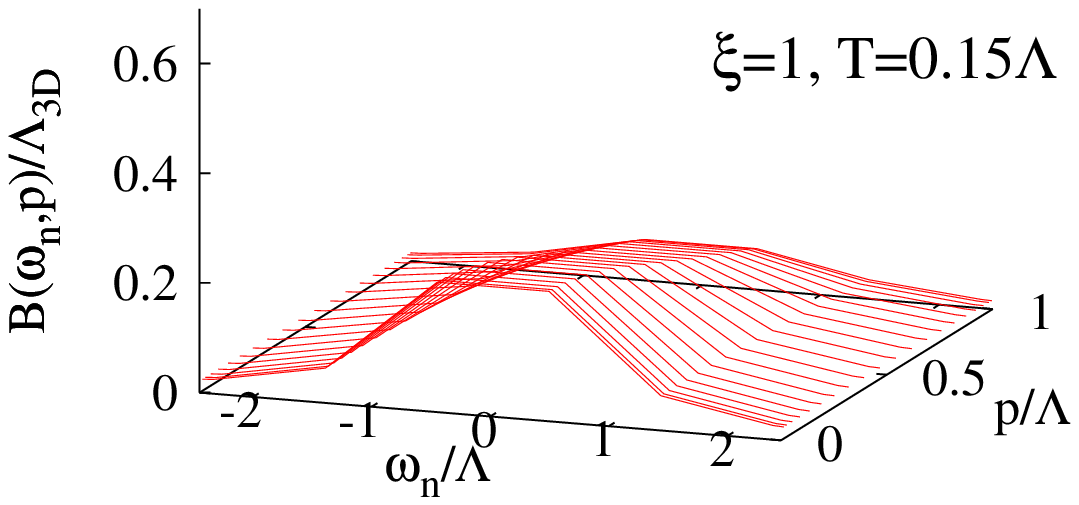}
  \includegraphics[width=5.3cm,keepaspectratio]{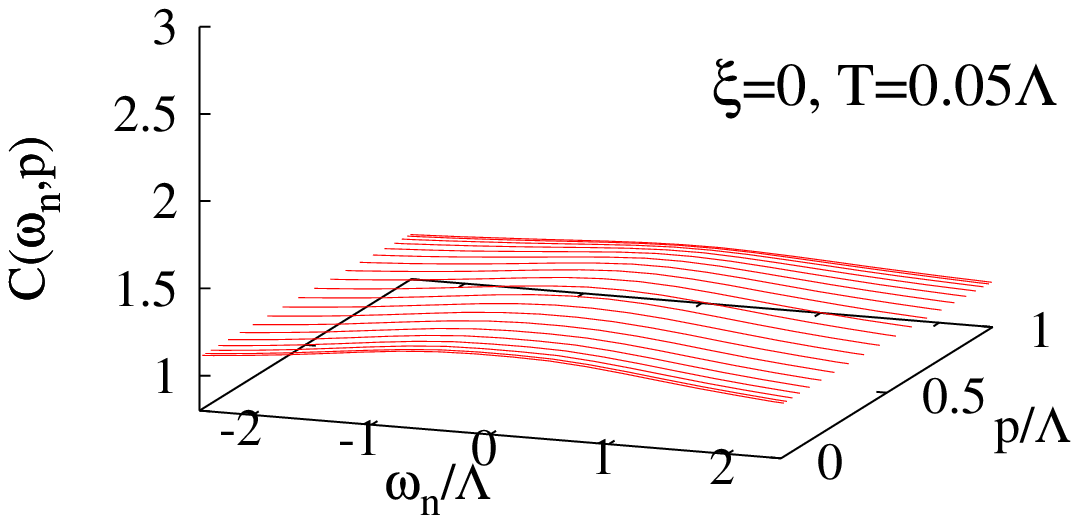}
  \includegraphics[width=5.3cm,keepaspectratio]{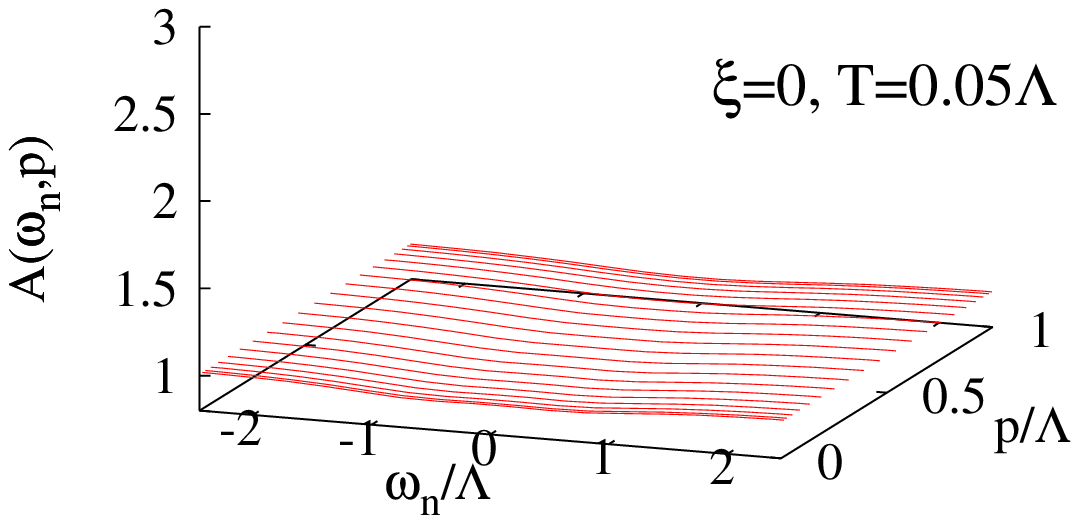}
  \includegraphics[width=5.3cm,keepaspectratio]{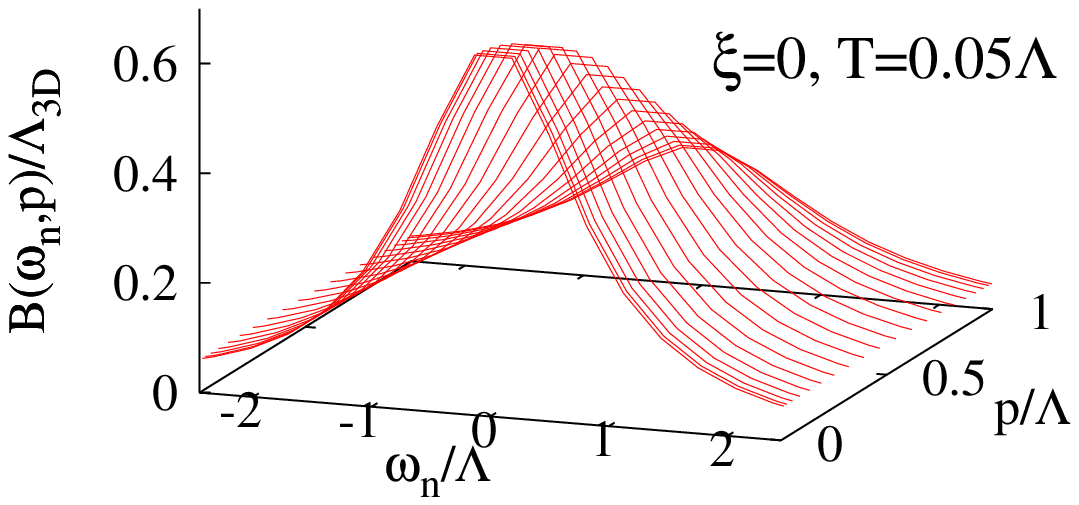}
  \includegraphics[width=5.3cm,keepaspectratio]{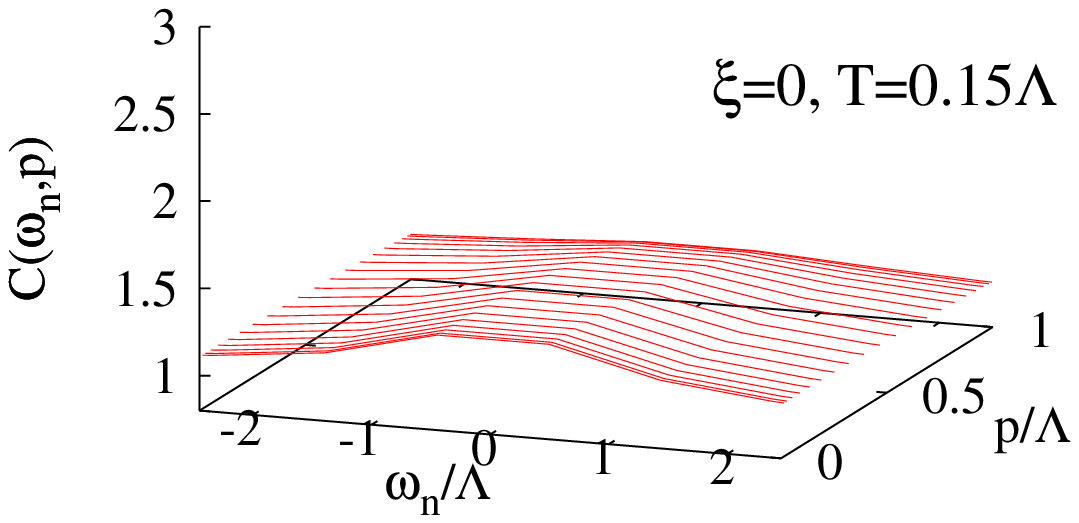}
  \includegraphics[width=5.3cm,keepaspectratio]{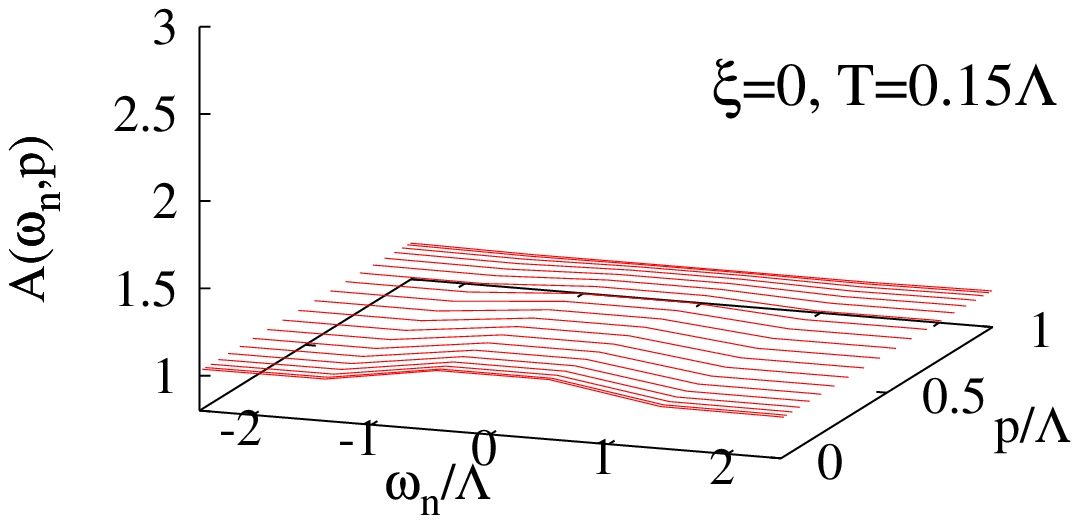}
  \includegraphics[width=5.3cm,keepaspectratio]{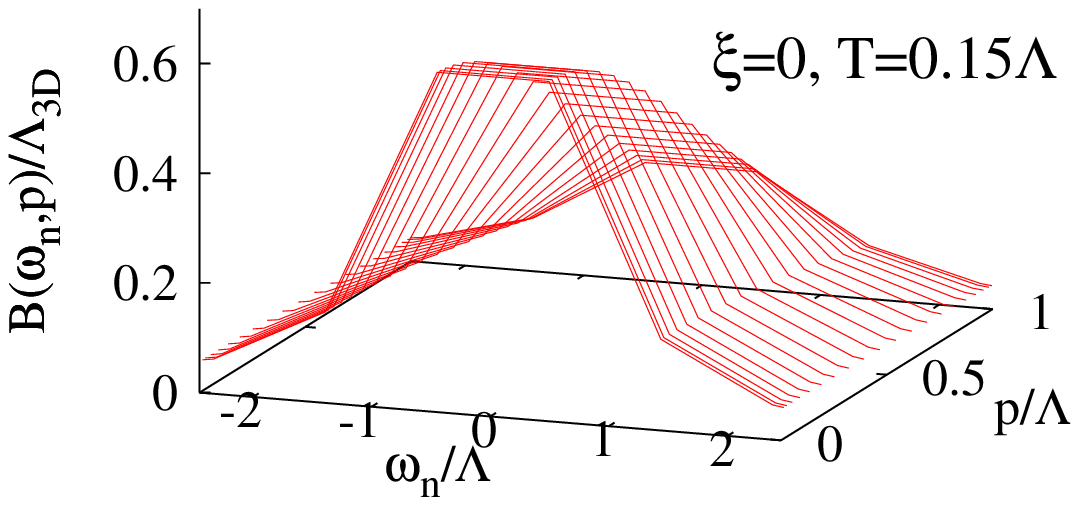}
  \caption{\label{fig:ABC}
    $C(\omega_n, p)$,  $A(\omega_n, p)$, $B(\omega_n, p)$ with
    $\alpha=3.5$, for $T=0.05 \Lambda$ and $T = 0.15\Lambda$.}
\end{center}
\end{figure}

Figure \ref{fig:ABC} displays the solutions of
$C(\omega_n,p)$, $A(\omega_n,p)$, $B(\omega_n,p)$ for $\alpha =3.5$,
$\lambda = 0.01\Lambda$, $m_0=0$ with $\xi =3$, $1$ and $0$ at 
temperatures $T=0.05\Lambda$ and $0.15\Lambda$. In carrying out the
calculations, we took the number of frequency
summation $N_t = 50$ and performed the $100$ times of iterations. We have
numerically confirmed that these numbers are good enough in our present
analyses, namely the solutions well converge with $N_t=50$ and $N_i=100$.
Then, we will persistently use this setting in what follows.

From Fig.~\ref{fig:ABC}, one sees that $C$, $A$ and $B$ decrease
with increasing temperature, and these shapes become smoother
when temperature increases. This result on temperature can be understood
easily, since the non-perturbative effect is expected to be suppressed
at high temperature. Therefore the symmetry tends to be restored at
high temperature, which can be confirmed by the obtained solutions.
On the gauge dependence, the values of $C$ and $A$ are large
for large $\xi$, while $B$ becomes smaller with respect to $\xi$.
It is known from the $T=0$ analyses with the four dimensional
momentum cutoff regularization that $C(=A)$ and $A$ become $1$
in the Landau gauge $\xi=0$. Then the deviation from $1$ is expected
to become larger when $\xi$ increases, which is confirmed by the
obtained results. On the other hand, the value of $B$ is small for large
$\xi$. This comes from the fact that the denominator of the fermion
propagator becomes larger when $C$ and $A$ raise up, then
the integral in the equation for $B$ becomes small, leading smaller
value for $B$ consequently.

\subsection{\label{subsec:t_dependence}
Temperature dependence on the solutions}
For the purpose to see the temperature dependence, it may be nice to
display the figures with the temperature axis.
\begin{figure}[!h]
\begin{center}
  \includegraphics[width=5.3cm,keepaspectratio]{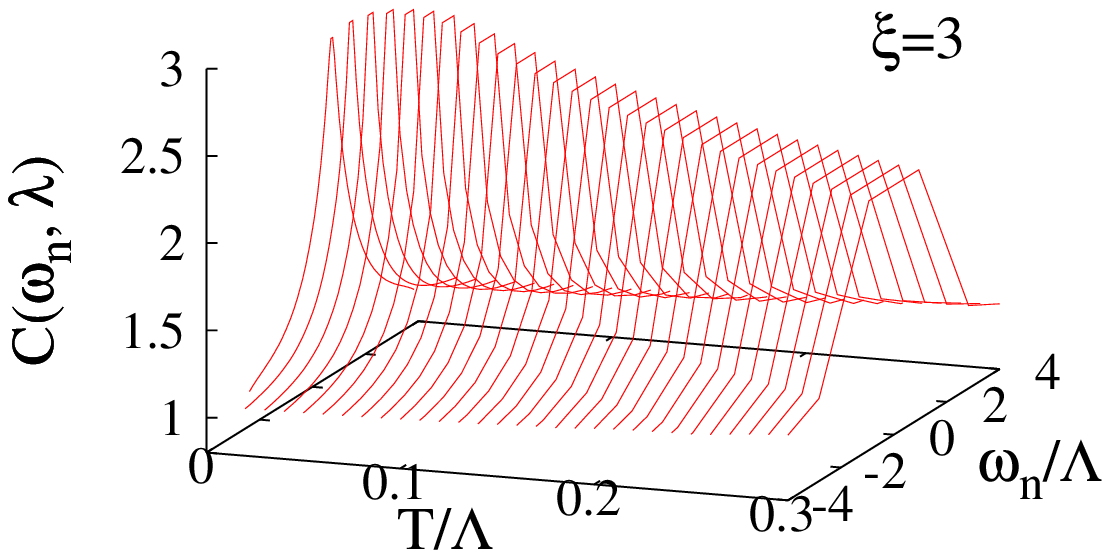}
  \includegraphics[width=5.3cm,keepaspectratio]{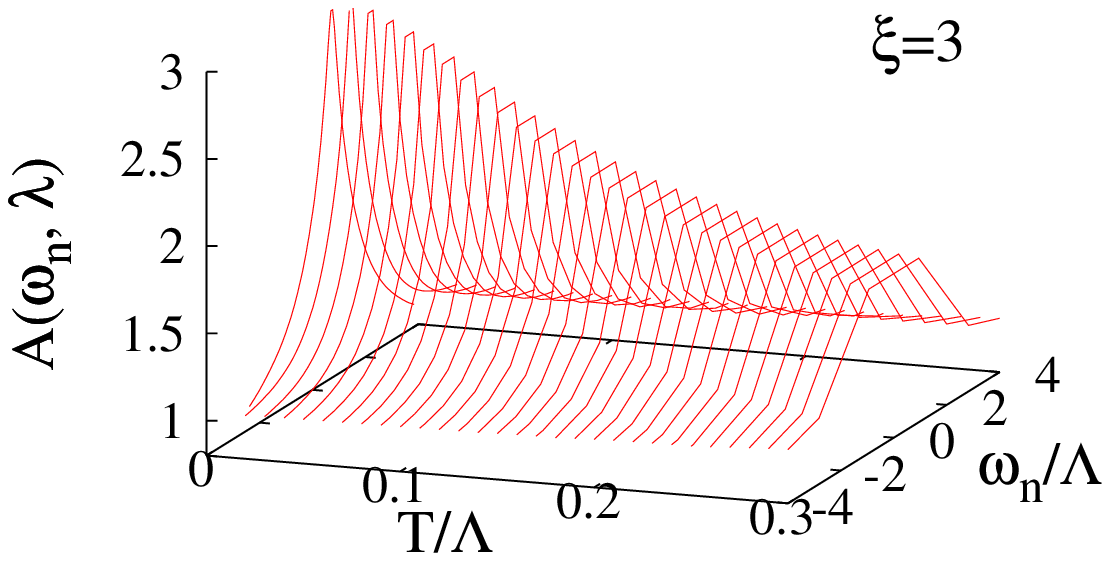}
  \includegraphics[width=5.3cm,keepaspectratio]{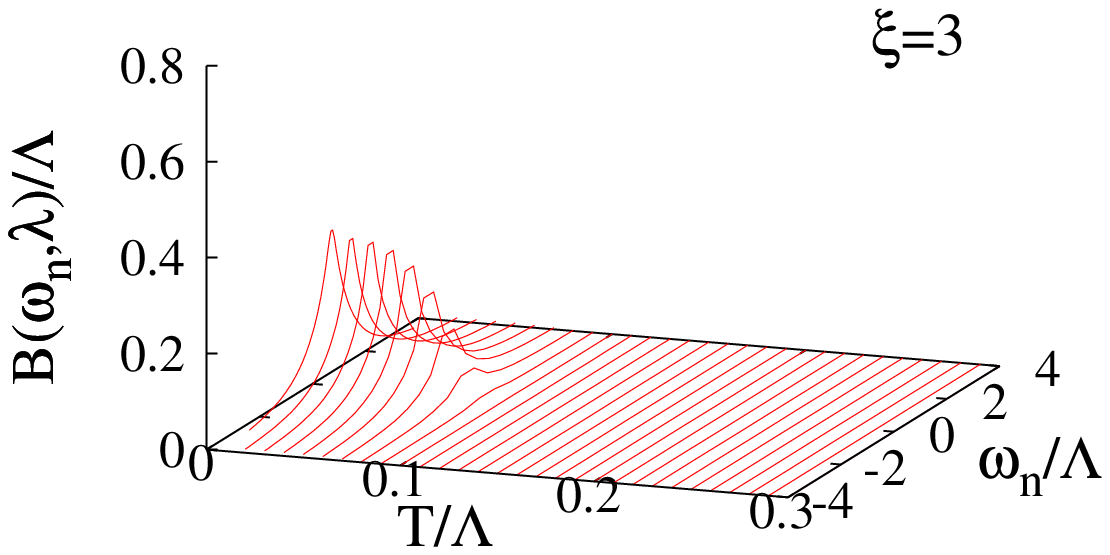}
  \includegraphics[width=5.3cm,keepaspectratio]{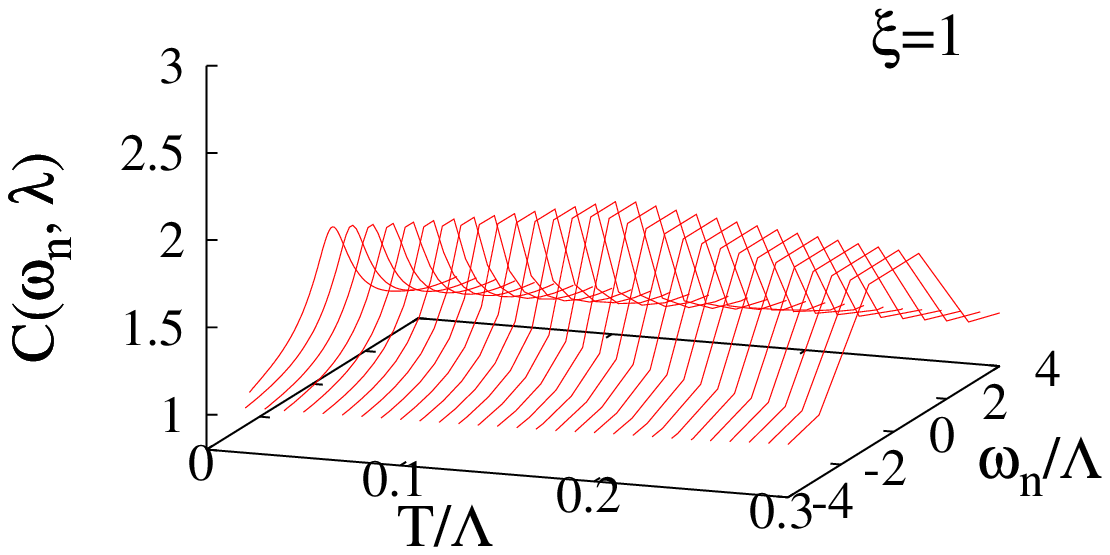}
  \includegraphics[width=5.3cm,keepaspectratio]{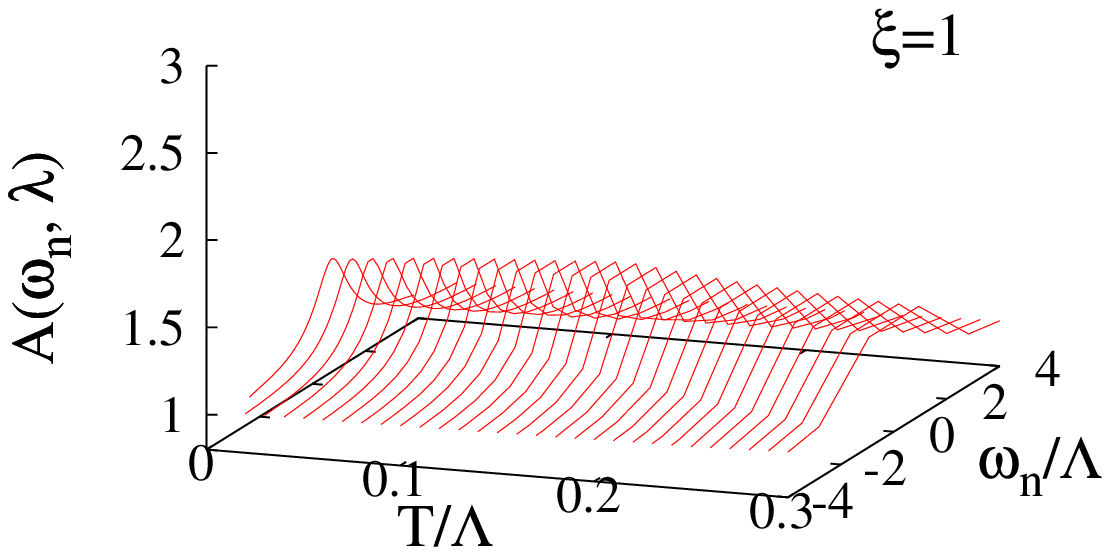}
  \includegraphics[width=5.3cm,keepaspectratio]{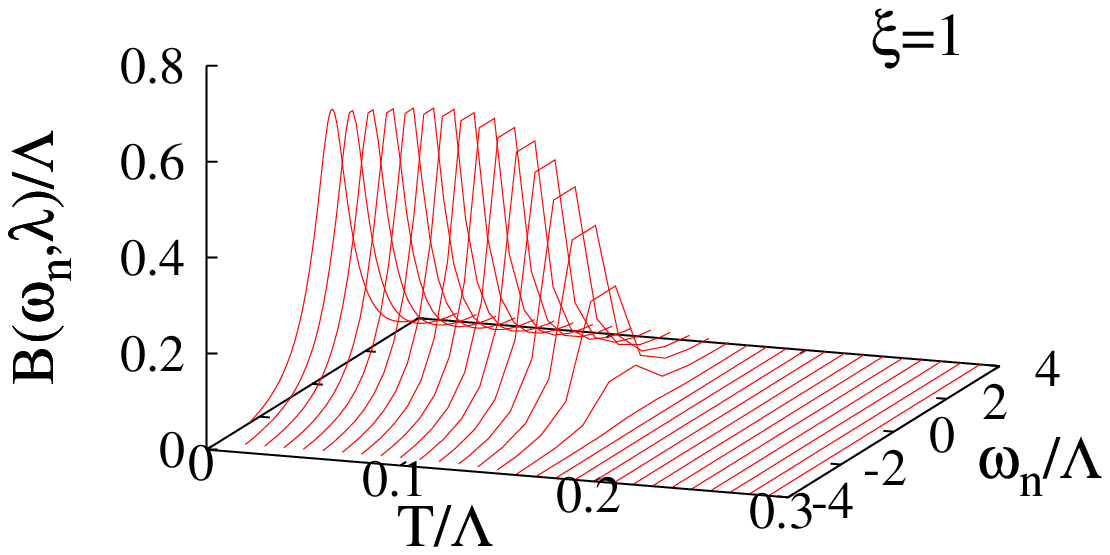}
  \includegraphics[width=5.3cm,keepaspectratio]{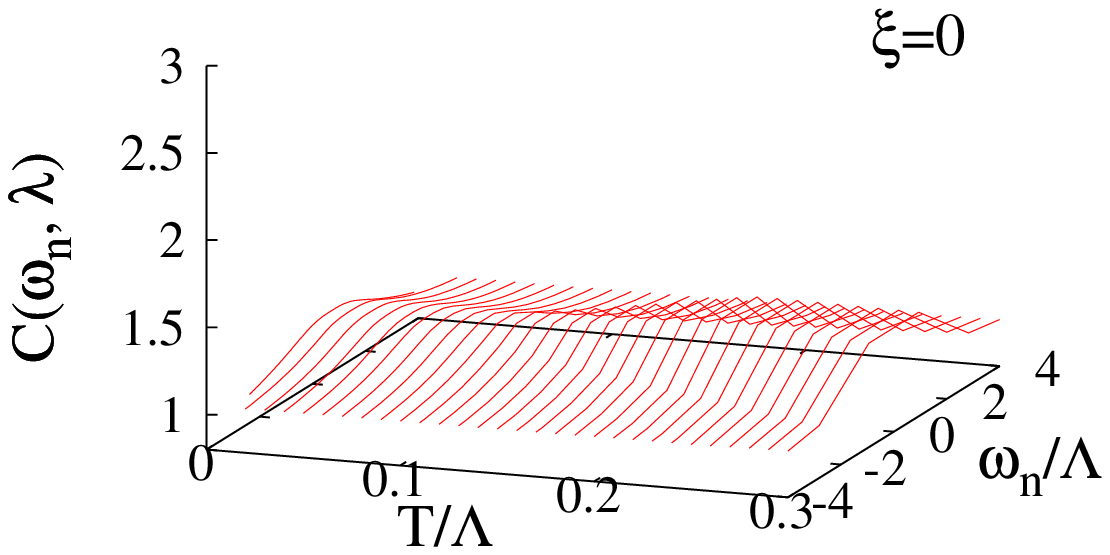}
  \includegraphics[width=5.3cm,keepaspectratio]{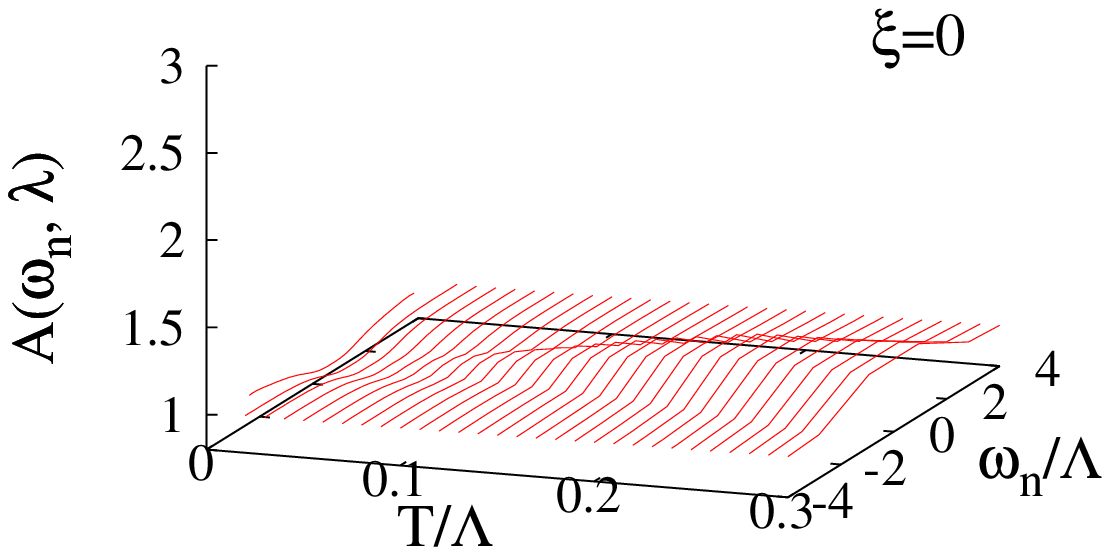}
  \includegraphics[width=5.3cm,keepaspectratio]{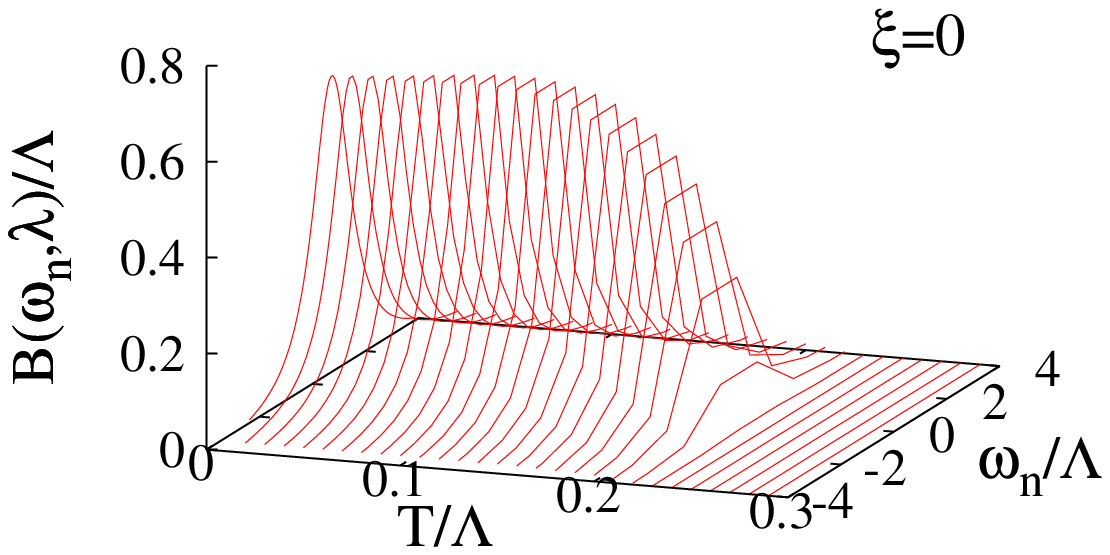}
  \caption{\label{fig:ABCtn}
    $C(\omega_n, \lambda)$,  $A(\omega_n, \lambda)$,
    $B(\omega_n, \lambda)$
    for $\alpha=3.5$.}
\end{center}
\end{figure}
\begin{figure}[!h]
\begin{center}
  \includegraphics[width=5.3cm,keepaspectratio]{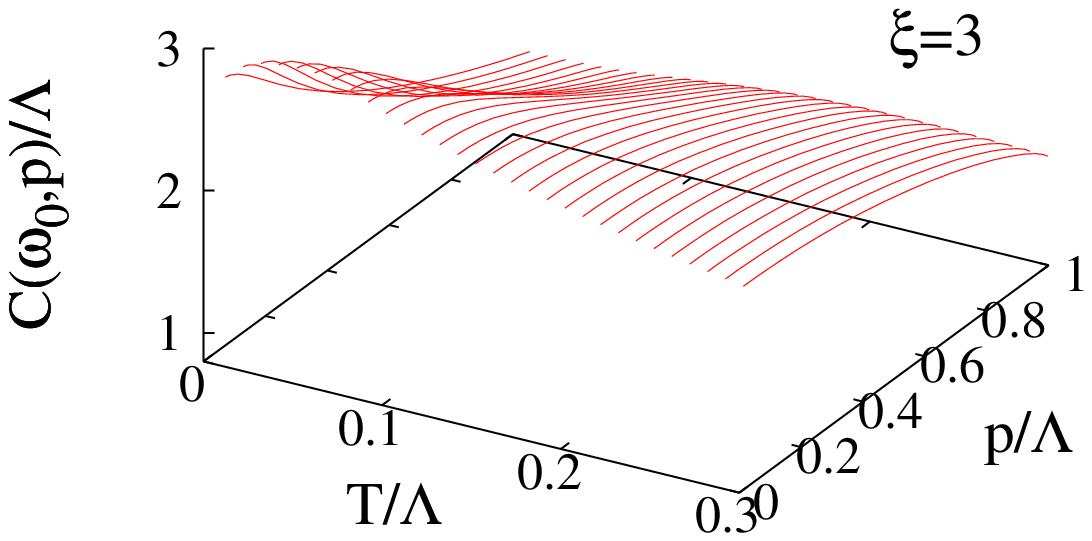}
  \includegraphics[width=5.3cm,keepaspectratio]{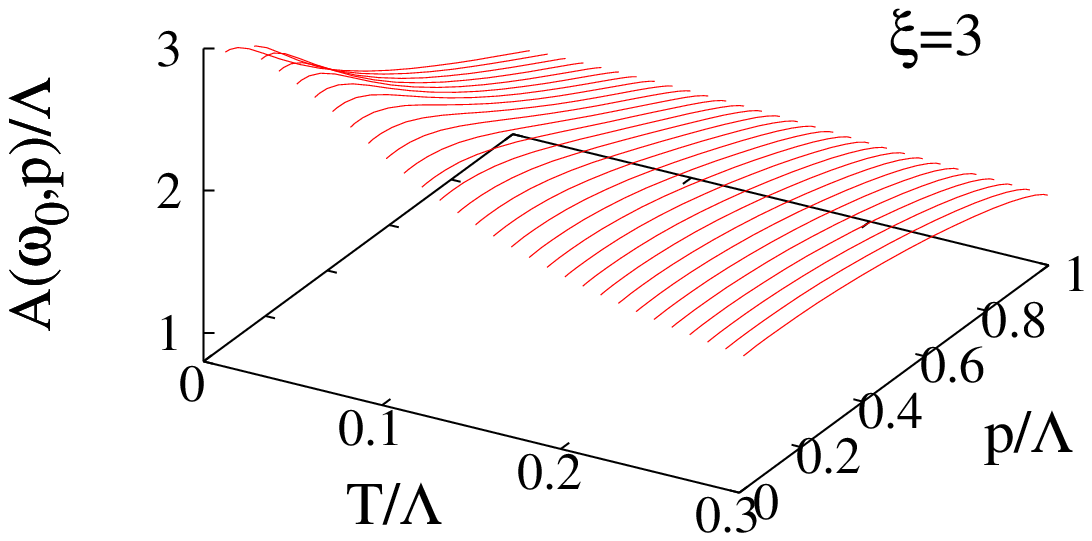}
  \includegraphics[width=5.3cm,keepaspectratio]{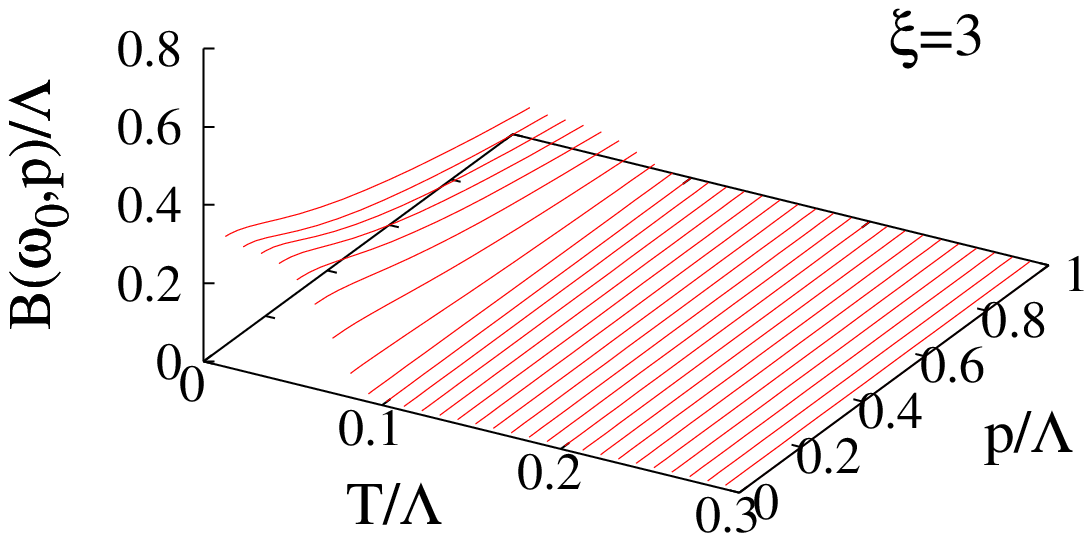}
  \includegraphics[width=5.3cm,keepaspectratio]{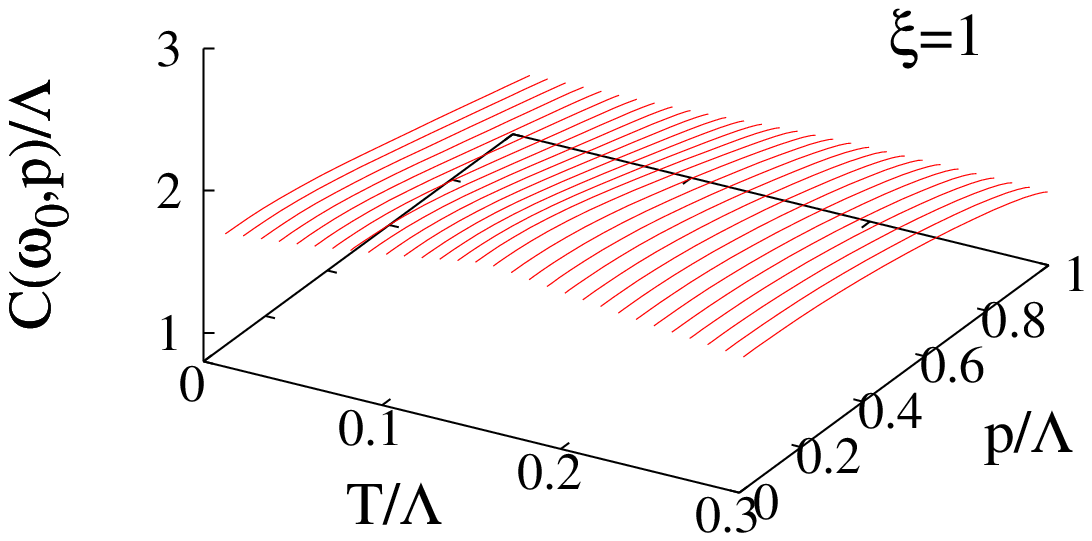}
  \includegraphics[width=5.3cm,keepaspectratio]{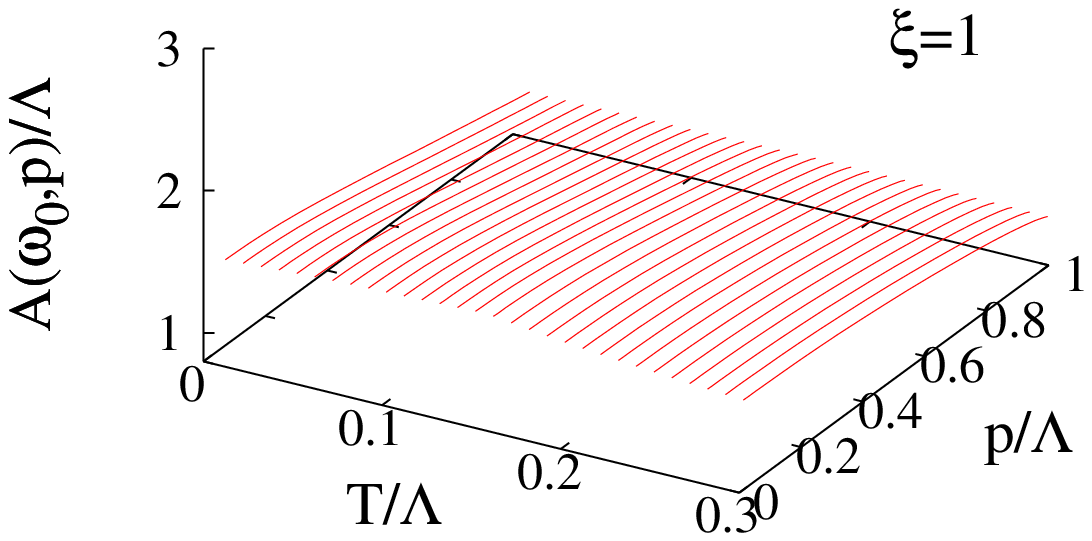}
  \includegraphics[width=5.3cm,keepaspectratio]{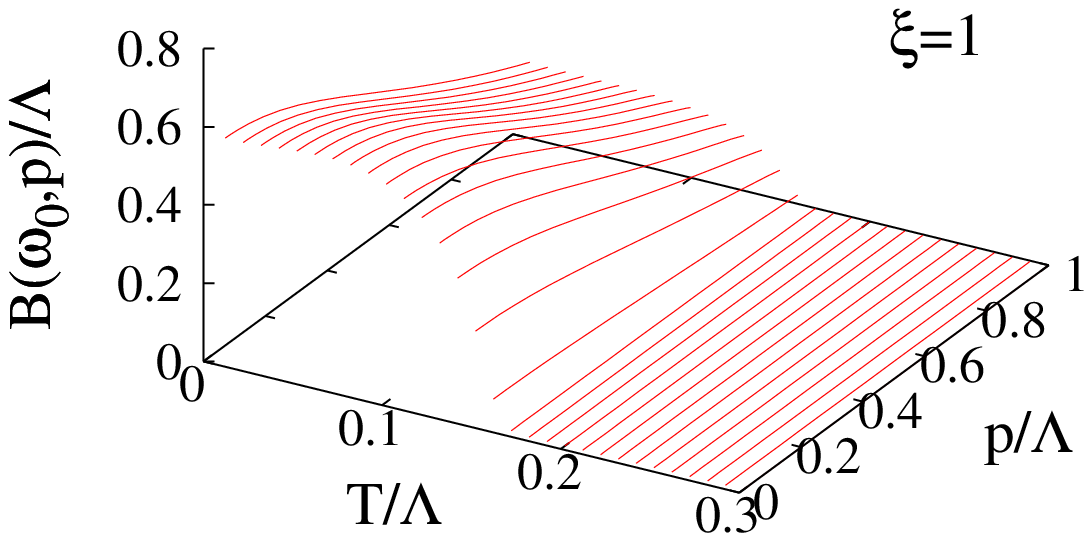}
  \includegraphics[width=5.3cm,keepaspectratio]{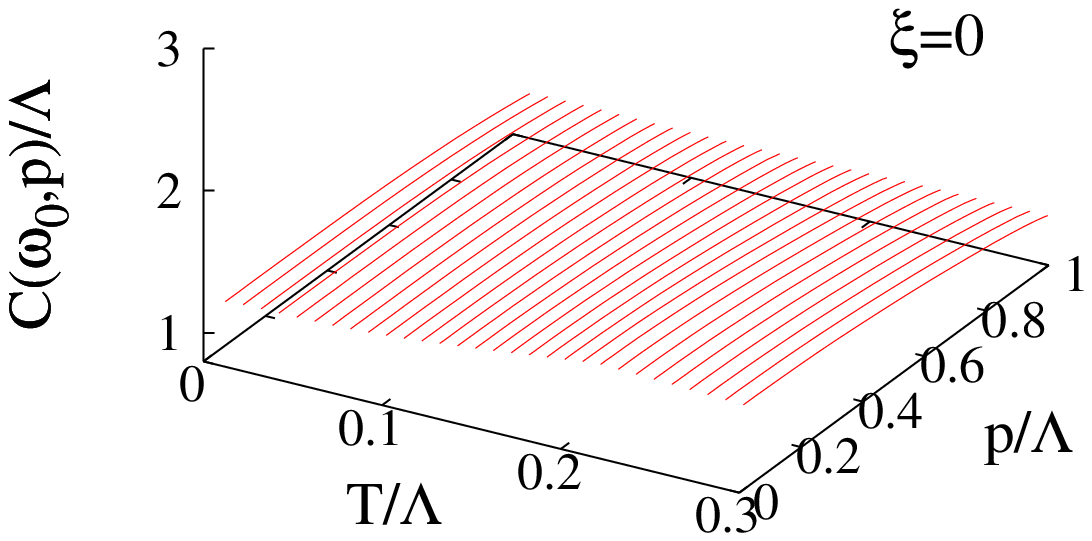}
  \includegraphics[width=5.3cm,keepaspectratio]{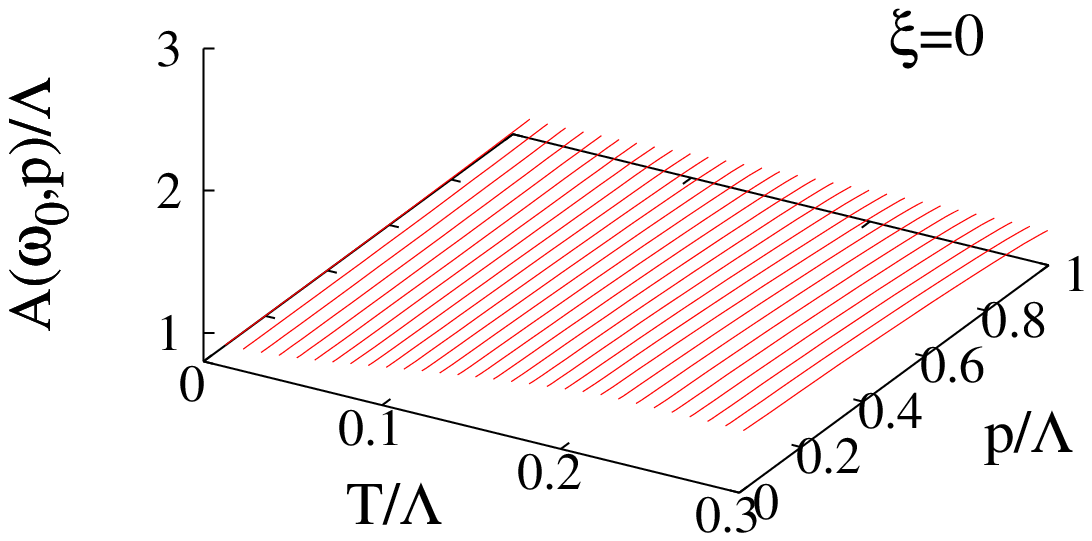}
  \includegraphics[width=5.3cm,keepaspectratio]{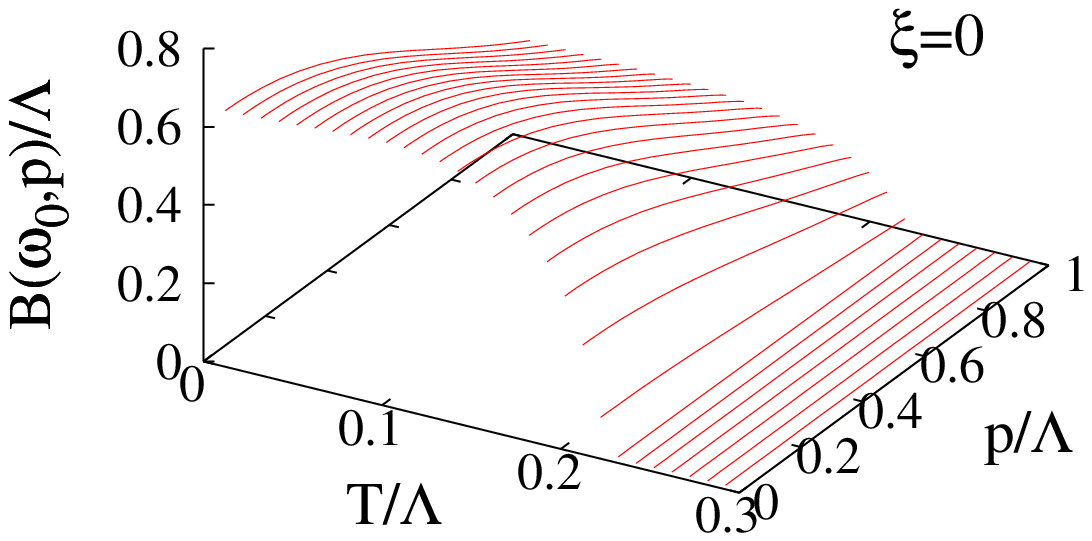}
  \caption{\label{fig:ABCtp}
    $C(\omega_0, p)$,  $A(\omega_0, p)$,
    $B(\omega_0, p)$
    for $\alpha=3.5$.}
\end{center}
\end{figure}
Figures~\ref{fig:ABCtn} and \ref{fig:ABCtp} show the temperature
dependence on $C(\omega_n, \lambda)$,  $A(\omega_n, \lambda)$,
 $B(\omega_n, \lambda)$, and  $C(\omega_0, p)$, $A(\omega_0, p)$,
 $B(\omega_0, p)$. We observe that $C$ and $A$ slightly increases
 according to $T$ when $B \neq 0$, then decreases after $B$ vanishes.
 While the result of $B$ is easier to see; it stays almost constant
 for low $T$, then drops  at certain $T$. This is the clear signal
 of the phase transition, where the mass factor $B$ becomes zero
 at some critical temperature $T_c$. The issue is interesting;
 we will discuss on this in more detail in the next section.

\section{\label{sec:re_cr}
Chiral phase transition}
We have shown rough shapes of the solutions by aligning the three
dimensional figures in the previous section. We will now discuss the
gauge dependence on the chiral phase transition in more detail focusing
on the solutions at the lowest frequency and momentum, i.e., the
values at $(\omega_n,p) = (\omega_0,\lambda)$.

\subsection{\label{subsec:gauge_dependence}
Gauge dependence on the solutions}
\begin{figure}[!h]
\begin{center}
  \includegraphics[width=7.0cm,keepaspectratio]{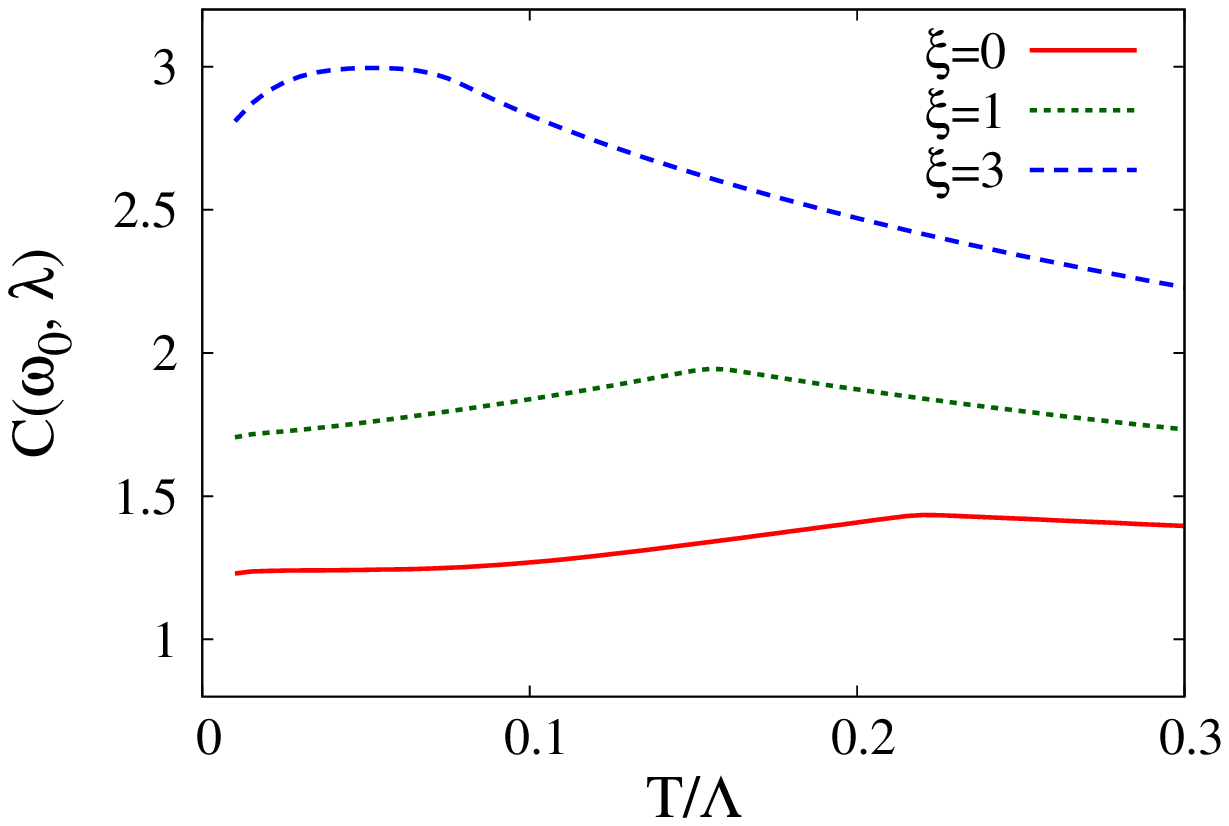}
  \includegraphics[width=7.0cm,keepaspectratio]{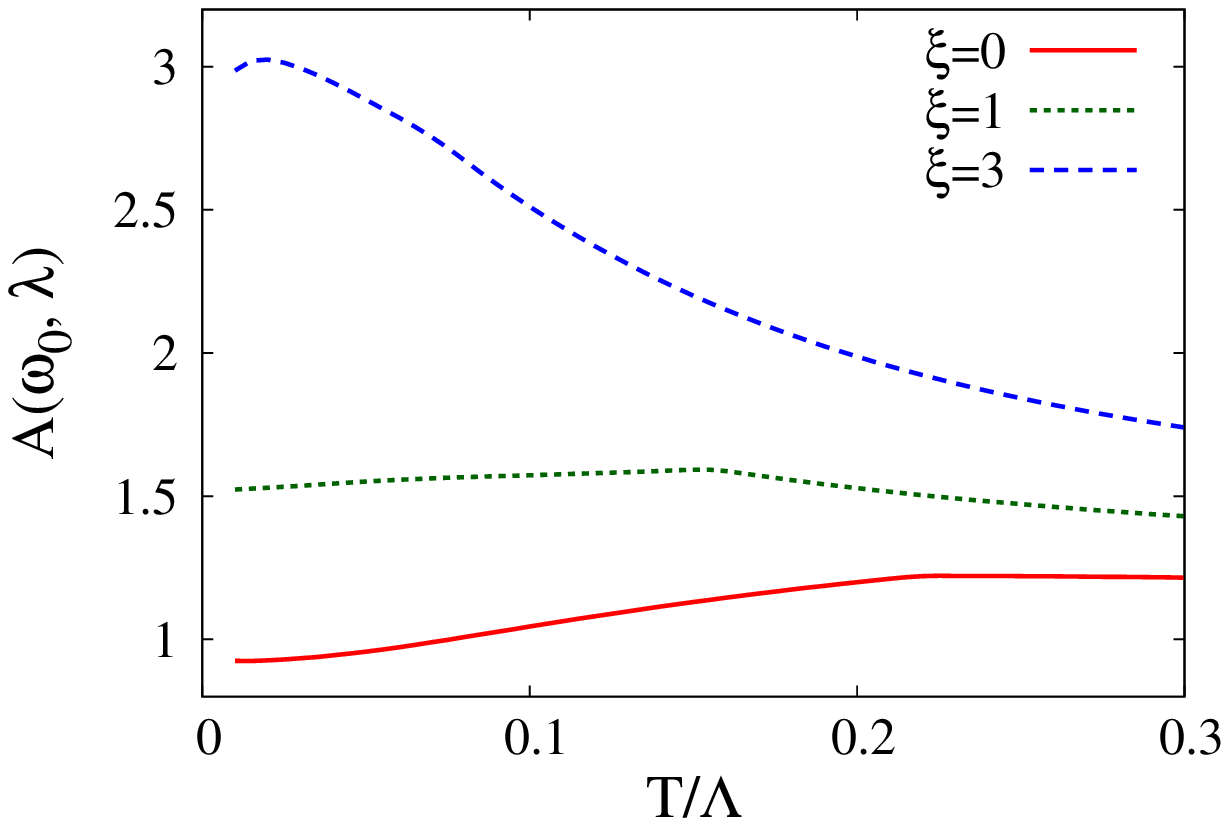}
  \includegraphics[width=7.0cm,keepaspectratio]{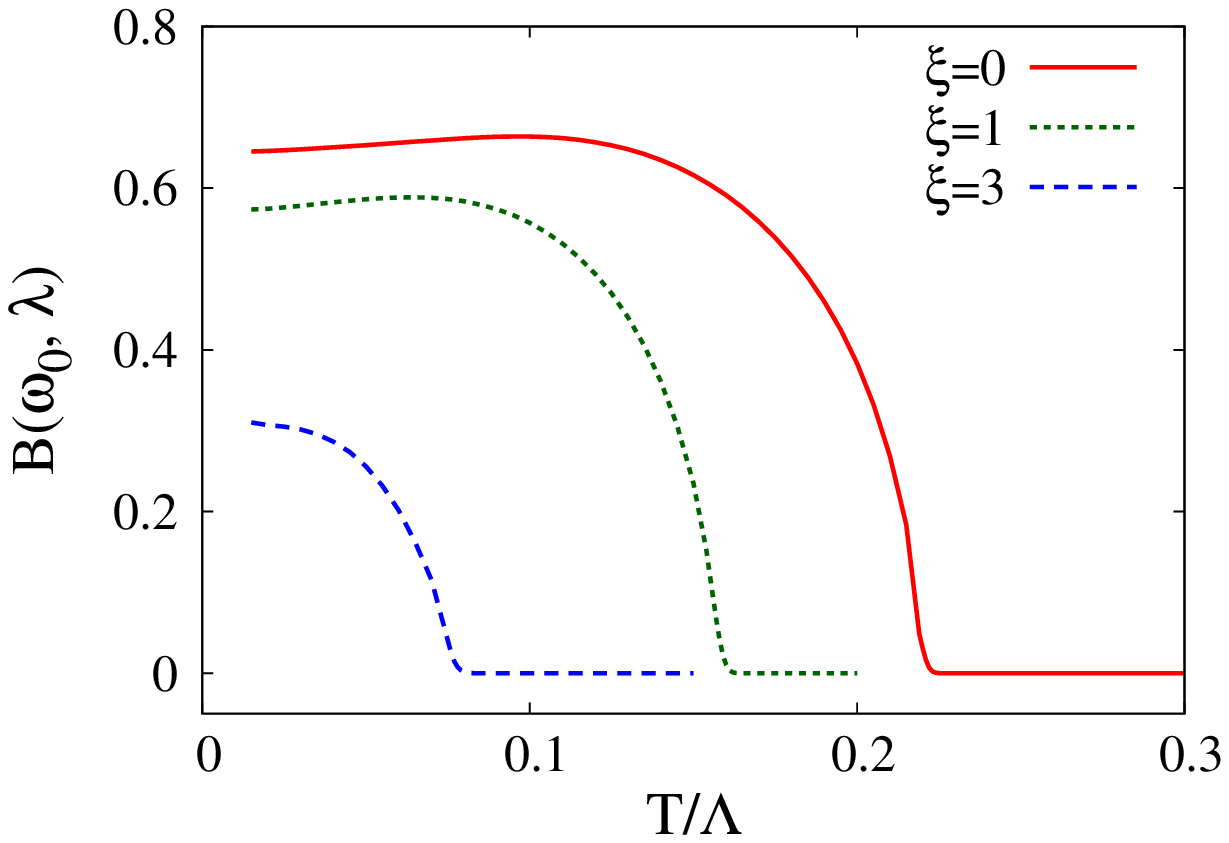}
  \includegraphics[width=7.0cm,keepaspectratio]{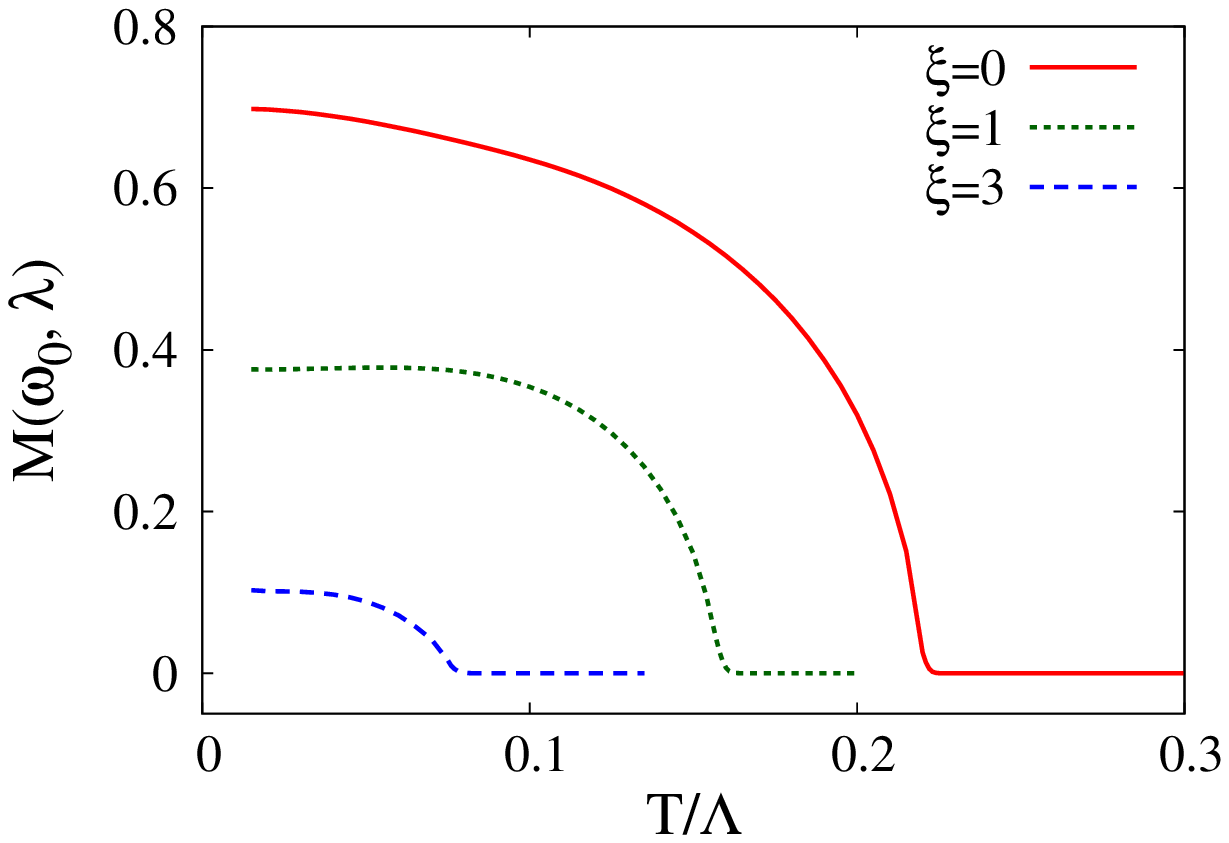}
  \caption{\label{fig:ABC00t}
  $C$, $A$, $B$ and $M$ as the
  functions of $T$ with $\alpha=3.5$ for $\xi=0$, $1$ and $3$ .}
\end{center}
\end{figure}

Figure \ref{fig:ABC00t} shows how the solutions $C(\omega_0,\lambda)$,
$A(\omega_0,\lambda)$, $B(\omega_0,\lambda)$ and
$M(\omega_0,\lambda)$ changes with respect to $T$.
As we previously saw that $C$ increases with $T$ when $B$ is nonzero,
then decreases after $B$ vanishes. $A$ shows the similar curves for
$\xi=0$ and $1$, while it almost monotonously decreases for $\xi=3$.
$B$ and $M$ become small for high $T$ as already observed in
Figs. \ref{fig:ABCtn} and \ref{fig:ABCtp}.

From the figure, one can clearly confirm the quantitative gauge dependence
on the solutions. $C$ and $A$ are around $3$ for low $T$ with $\xi=3$,
and within the range of around $1-2$ for $\xi=0$ and $1$.  $M$ at $T=0$
is around $0.7$ for $\xi=0$ and it becomes considerably smaller for
$\xi=3$. Thus, there exists the indispensable gauge dependence on
the solutions for the quenched SDE.

\subsection{\label{subsec:critical}
Critical temperature}
Let us study the critical temperature, $T_c$, of the chiral phase
transition here. We have discussed that the dynamically generated
mass can be regarded as the order parameter of the transition in
Sec.~\ref{subsec:equation}. Then we shall carefully search the critical
temperature of the chiral phase transition by evaluating the value
of $M$.

\begin{figure}[!h]
\begin{center}
  \includegraphics[width=7.0cm,keepaspectratio]{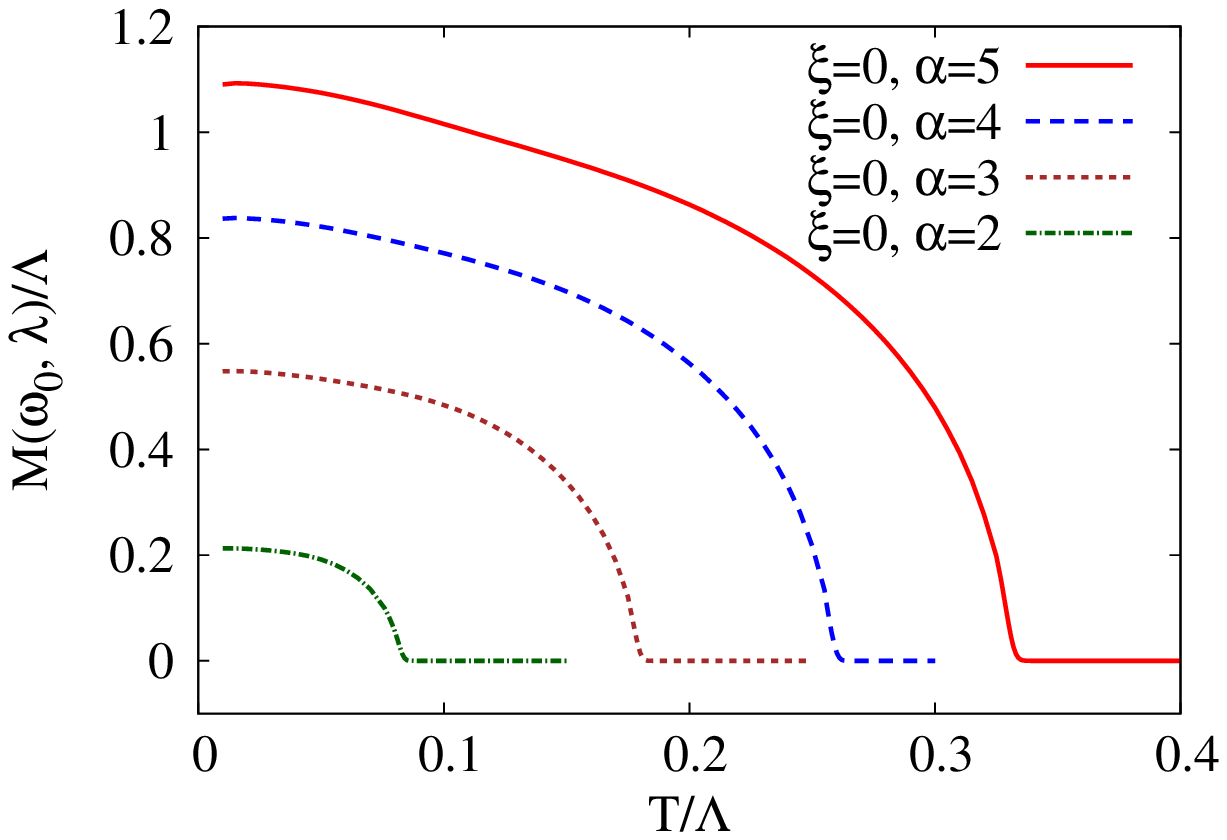}
  \includegraphics[width=7.0cm,keepaspectratio]{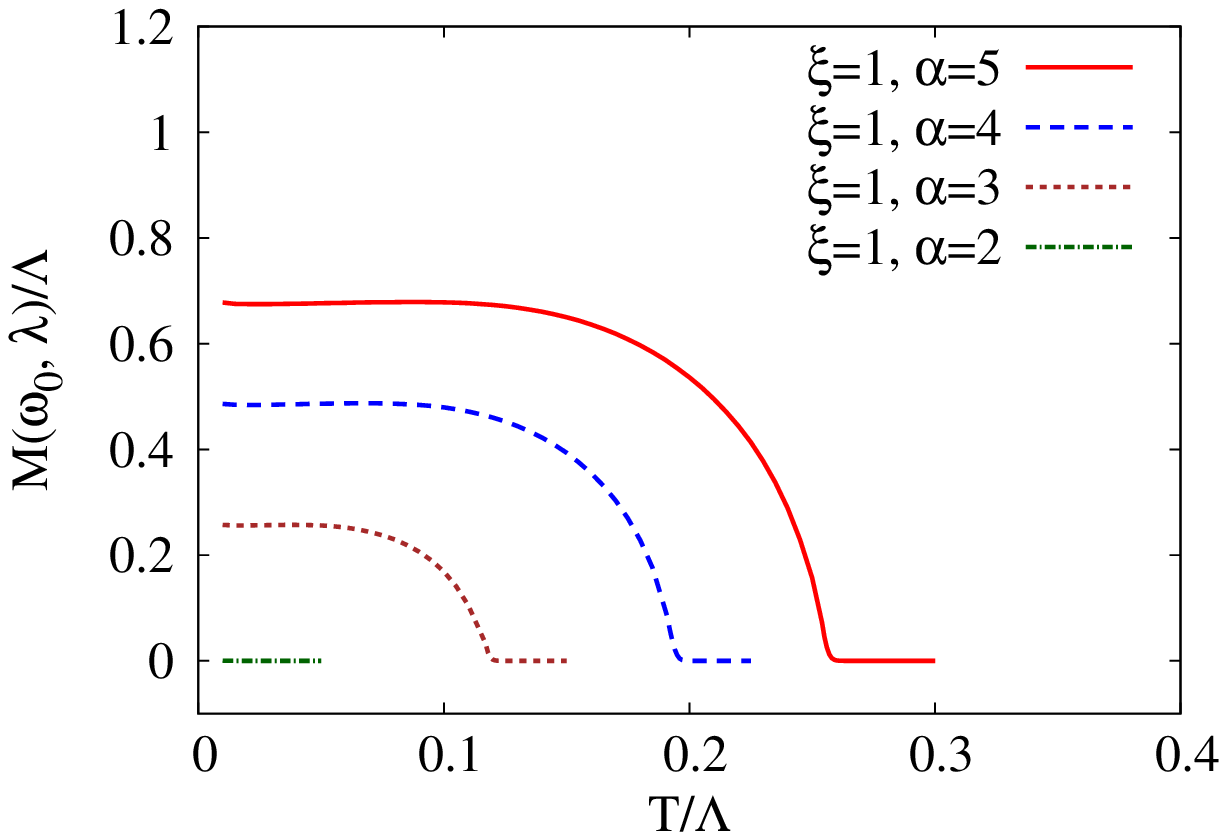}
  \includegraphics[width=7.0cm,keepaspectratio]{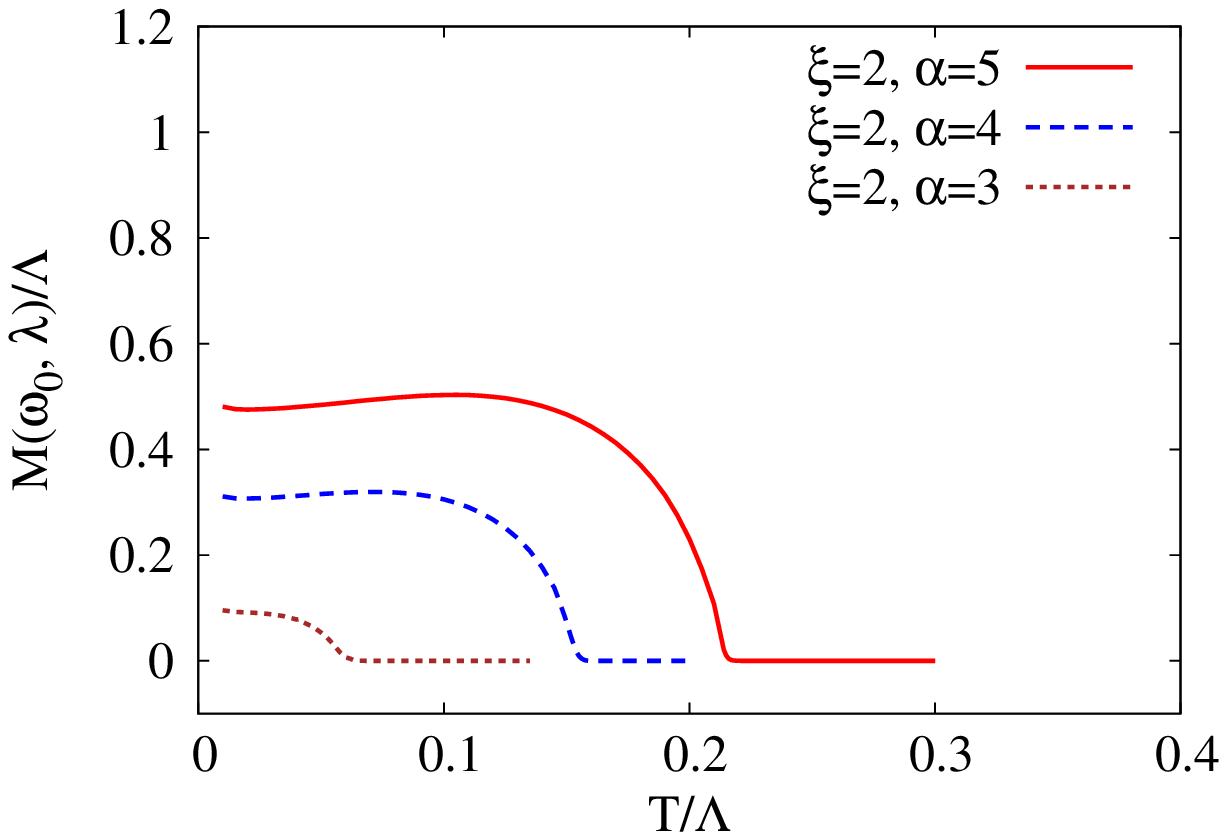}
  \includegraphics[width=7.0cm,keepaspectratio]{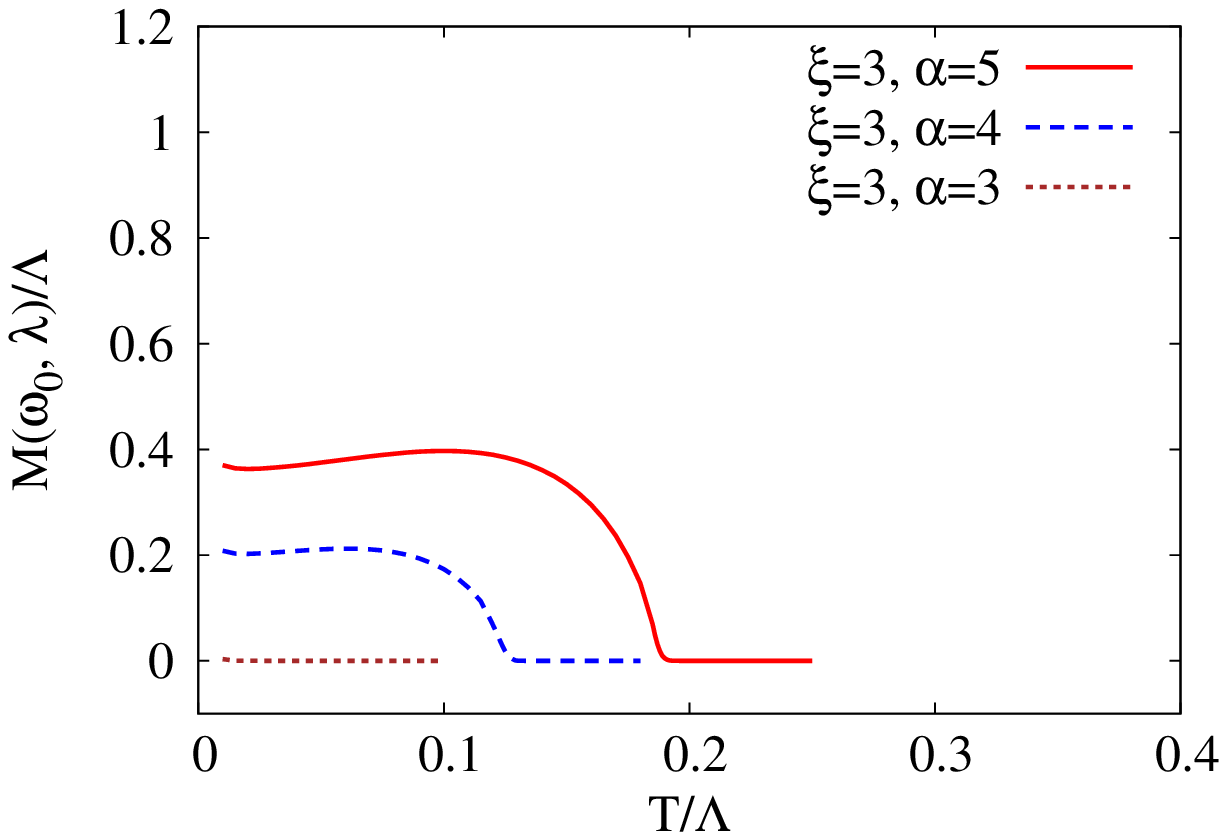}
  \caption{\label{fig:Mt}
  $M(\omega_0, \lambda)$ as the
  functions of $T$ for various $\xi$ and $\alpha$.}
\end{center}
\end{figure}
Figure \ref{fig:Mt} displays the temperature dependence on the
dynamical mass, $M(\omega_0,\lambda)$, for various $\xi$ and
$\alpha$. The critical temperature becomes larger when $\alpha$
increases. This can be easily understood since the system has
stronger correlation for larger coupling. While $T_c$ is smaller
for larger $\xi$. This comes from the effect of the field renormalization
as we have seen from the preceding section. The explicit numbers
of the critical temperature for various $\xi$ and $\alpha$ are aligned
in Tab. \ref{tab:Tc}.
\begin{table}[h!]
\caption{Critical temperature ($T_c/\Lambda$).}
\label{tab:Tc}
\begin{center}
\begin{tabular*}{10cm}{@{\extracolsep{\fill}}ccccc}
\hline
   & $\alpha=2$ & $\alpha=3$ & $\alpha=4$  & $\alpha=5$ \\
\hline
$\xi=0$ & $0.089$ & $0.187$ & $0.267$ & $0.341$ \\
$\xi=1$ & --- & $0.126$ & $0.203$  & $0.265$  \\
$\xi=2$ & --- & $0.07$   & $0.162$  & $0.223$  \\
$\xi=3$ & --- & --- & $0.14$  & $0.197$ \\
\hline
\end{tabular*}
\end{center}
\end{table}

\begin{figure}[!h]
\begin{center}
  \includegraphics[width=7.0cm,keepaspectratio]{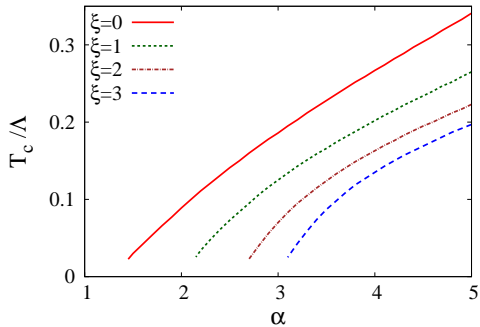}
  \caption{\label{fig:Tc_a}
   Critical temperature with respect to $\alpha$.
   }
\end{center}
\end{figure}

Finally, we show $T_c$ as the function of $\alpha$ in Fig. \ref{fig:Tc_a}.
We again confirm that there exists substantial gauge dependence on the
critical behavior. Note that we can not extend the curves for very
low temperature within this formalism, since we need to take large number
of frequency summation to get reliable results for low
$T_c$~\cite{Chen:2009mv}, which requires enormous calculation time.
However one easily sees the limiting values on the critical coupling for
$T_c \to 0$ from the figure, which are around
$\alpha_c \simeq 1.2$, $2$, $2.6$ and $3$ for $\xi=0$, $1$, $2$ and $3$.
It may be interesting to mention that these values are close to the ones
obtained in the SDE with the four dimensional cutoff regularization at
$T=0$, which leads $\alpha_c \simeq \pi/3$,
$2$ and $3$ for $\xi=0$, $1$ and $3$~\cite{Kizilersu:2014ela}.

\section{\label{sec:conclusion}
Concluding remarks}
We have evaluated the numerical solutions for the SDE at finite temperature
without approximations in Abelian gauge theory in this paper. There we
observe the considerable gauge dependence on the solutions. We also
analyzed the critical behavior of the chiral phase transition within the SDE
approach. We confirm that the critical temperature as well drastically
depends on the chosen gauge. This indicates that, when one investigates
the chiral phase transition in the quenched SDE with using some
approximations, one should be very careful on the set of assumptions.
In particular, the field renormalization factors $C$ and $A$ nonnegligibly
deviate from the unity for nonzero $\xi$.

Since we started from the SDE with the quenched form, then the gauge
dependence as observed in this analysis is inevitable. This is clearly
unsatisfactory feature of the quenched equations. Therefore the further
analyses with generalized equations,  such as the unquenched
investigations \cite{Kizilersu:2014ela},
or the extended vertex \cite{Ball:1980ay, Curtis:1990zs, Kizilersu:2009kg} 
and coupling \cite{Roberts:1989mj},  are obviously needed; this
should be the future direction in the analyses based on the
Schwinger--Dyson equation.

\begin{acknowledgments}
The author thanks to T. Inagaki and H. Mineo for discussions.
The author is supported by Ministry of Science and Technology
(Taiwan, ROC), through Grant No. MOST 103-2811-M-002-087.
\end{acknowledgments}


\end{document}